\documentclass[letterpaper,twocolumn,english,aps,citeautoscript,superscriptaddress,longbibliography]{revtex4-1}
\usepackage[T1]{fontenc}
\usepackage[latin9]{inputenc}
\usepackage{amsmath}
\usepackage{amssymb}
\usepackage{graphicx}

\makeatletter

\pdfpageheight\paperheight
\pdfpagewidth\paperwidth

\usepackage{hyperref}
\hypersetup{colorlinks=true, citecolor=black, hidelinks = true}
\usepackage{amsfonts}
\usepackage{bbm}
\usepackage{graphics}

\def\ket#1{\left| #1\right>}
\def\bra#1{\left< #1\right|}

\def\<{\langle}
\def\>{\rangle}

\usepackage{leftidx}

\usepackage{etoolbox}
\patchcmd{\subsection}{\centering}{\raggedright}{}{}

\makeatother

\usepackage{babel}
\begin{document}
\title{Enabling Modularity for Spin Qubits via Driven Quantum Dot-Mediated
Entanglement}
\author{V. Srinivasa}
\email{vsriniv@uri.edu}

\affiliation{Department of Physics, University of Rhode Island, Kingston, RI 02881,
USA}
\date{\today}
\begin{abstract}
We present an approach for entangling spin qubits via capacitive coupling
mediated by an ac electric field-driven multielectron mediator quantum
dot. To illustrate this method, we consider the case of a driven two-electron
dot that mediates entanglement between resonant exchange qubits defined
in three-electron triple quantum dots, which enable direct capacitive
coupling and interaction with microwave fields via intrinsic spin-charge
mixing. The method can also be applied to other types of spin qubits
that can be coupled capacitively. We show that this approach leads
to rapid, single-pulse universal entangling gates for resonant exchange
qubits that are activated via the drive on the mediator dot. Unlike
conventional tunneling-based two-qubit gates between exchange-only
qubits, the capacitive interaction-based gates we describe do not
require an extensive sequence of pulses to mitigate leakage. We describe
how this drive-activated local entangling approach can be integrated
with the driven sideband-based long-range approach for cavity-mediated
entangling gates developed in our previous work in order to enable
modularity for spin-based quantum information processing. 
\end{abstract}
\maketitle

\section{\label{sec:Introduction}Introduction}

Modularity \cite{Taylor2005,Jiang2007,Monroe2014,Vandersypen2017,Tosi2017,Jnane2022}
represents a promising approach to scaling quantum information processing
platforms, in which few-qubit modules with locally optimized control
and entanglement serve as building blocks that are linked via robust
long-range interactions. The realization of a fully modular system
requires the coherent integration of entanglement mechanisms over
a wide range of distances. In platforms based on semiconductor spin
qubits \cite{Loss1998,Kane1998,Hanson2007RMP,Zwanenburg2013,Scappucci2021,Chatterjee2021,Burkard2023},
the exchange interaction enables rapid and coherent gates \cite{Loss1998,Kane1998,DiVincenzo2000Nature,Levy2002,Petta2005,He2019,Hendrickx2020,Noiri2022,Xue2022,Mills2022,Weinstein2023,WangCA2024,Madzik2025}
but has an inherently short range limited by the wave function overlap
of neighboring electrons \cite{Burkard1999,Lidar2000}. Scaling by
adding many spin qubits to a single array thus typically involves
a high density of associated control electronics and is challenging,
while modularity makes use of existing smaller, well-controlled qubit
arrays \cite{Beil2014thesis,Fedele2021,Kiczynski2022,Wang2022,Philips2022,Borsoi2024,WangCA2024,Zhang2025,George2025,Ha2025,John2025,Edlbauer2025,deFuentes2026}
and provides greater space for control via spatially distributed entanglement.
Investigations and demonstrations of entangling interaction mechanisms
for spin qubits with increased spatial range have accordingly been
carried out, including those that effectively extend the range of
exchange coupling itself such as quantum dot mediators \cite{Craig2004,Busl2013,Braakman2013,Mehl2014,Srinivasa2015,Stano2015,Baart2017,Malinowski2018,Croot2018,Malinowski2019,Cai2019,Deng2020,Fedele2021,Wang2023,Oxtoa2025arxiv,Duan2025arxiv},
spin chains \cite{Friesen2007,Srinivasa2007,Oh2010,Mohiyaddin2016,Kandel2019,Qiao2020,Qiao2021,Munia2024},
and spin shuttling \cite{Taylor2005,Fujita2017,Mills2019,Seidler2022,Noiri2022NComm,Langrock2023,Zwerver2023,Xue2024,vanRiggelenDoelman2024,Struck2024,DeSmet2025,White2026,Nemeth2026,Undseth2026arxiv}. 

Alternatively, spin qubits can be entangled via the intrinsically
longer-range Coulomb interaction, including capacitive coupling \cite{Taylor2005,Stepanenko2007,vanWeperen2011,Trifunovic2012,Shulman2012,Srinivasa2015PRB,Nichol2017,Rao2026arxiv}
as well as long-distance interactions enabled by coupling spins to
the electric field of photons in a microwave cavity using the approach
of circuit quantum electrodynamics \cite{Childress2004,Blais2004,Wallraff2004,Blais2007,Majer2007,Sillanpaa2007,Blais2021,Clerk2020,Burkard2020,Burkard2023}.
Recent demonstrations of strong spin-photon coupling \cite{Mi2018,Samkharadze2018,Landig2018,Landig2019,Yu2023,Ungerer2024,Jiang2025}
and coherent photon-mediated interaction of two single-electron silicon
spin qubits \cite{Borjans2020,Harvey-Collard2022,Dijkema2025} suggest
a promising path toward modular scaling of spin qubits via long-distance
distribution of entanglement on the scale of microwave wavelengths
($\sim1\ {\rm mm}-1\ {\rm cm}$). These Coulomb interaction-based
entangling approaches require the coupling of spin qubits to electric
fields. 

Rapid and universal all-electrical manipulation via control of exchange
interactions alone can be achieved for qubits encoded in three-electron
spin states \cite{DiVincenzo2000Nature,Meier2003PRL,Meier2003PRB,Laird2010,Gaudreau2011,Medford2013NNano,Medford2013,Taylor2013,Doherty2013,Shi2012,Kim2014,Eng2015,Shim2016,Russ2017,Andrews2019,Weinstein2023,Acuna2024,Stastny2025,Madzik2025,Broz2025arxiv,Bosco2026,Broz2026arxiv},
in contrast to single-spin or two-spin qubits for which universal
electrical control requires an additional spin-charge mixing mechanism
\cite{Srinivasa2013} such as spin-orbit coupling or spin-position
coupling via magnetic field gradients \cite{Hanson2007RMP,Burkard2023}.
Exchange-only qubits also allow for operation within decoherence-free
subspaces that protect against collective decoherence \cite{Lidar2000}
and specific encodings, including four-electron variants, that enable
suppression of leakage via an energy gap \cite{Meier2003PRL,Meier2003PRB,Weinstein2005,Srinivasa2009,Medford2013,Taylor2013,Doherty2013,Wardrop2016,Shim2016,Acuna2024,Broz2025arxiv,Bosco2026,Broz2026arxiv,Sala2017,Russ2018,Foulk2025}.
Nevertheless, universal two-qubit gates for conventional exchange-only
qubits remain challenging to implement as they require extensive pulse
sequences in order to mitigate leakage that exists even in the absence
of noise \cite{DiVincenzo2000Nature,Fong2011,Setiawan2014,Weinstein2023,Heinz2025,Madzik2025}.
This leakage results from the spin-conserving tunneling mechanism
underlying the exchange interaction, which conserves the total spin
of the two-qubit system but not that of individual qubits during gate
operation. 

The resonant exchange (RX) qubit \cite{Medford2013,Taylor2013} is
a particular form of the exchange-only qubit defined in a triple quantum
dot that enables high-frequency universal operation via resonant microwave
driving of the exchange at a symmetric operation point with suppressed
sensitivity of the qubit to low-frequency charge noise. While neighboring
RX qubits can be entangled via exchange with energetic suppression
of leakage \cite{Doherty2013,Wardrop2016}, the intrinsic spin-charge
coupling present in the logical RX qubit states also enables direct
capacitive interaction \cite{Taylor2013,Pal2014,Pal2015,Feng2021}
as well as entanglement over long distances using microwave cavity
photons via multiple approaches \cite{Taylor2013,Russ2015b,Srinivasa2016}.
In contrast to exchange, these Coulomb interaction-based mechanisms
limit leakage by conserving the spin of the individual RX qubits.
As we have shown in recent theoretical work \cite{Srinivasa2024},
parametrically driven RX qubits coupled via microwave photons in a
cavity can be entangled using sideband resonances even with mutually
off-resonant qubit and cavity frequencies. The drive-tunable nature,
spectral flexibility, and suppressed sensitivity to cavity photon
decay that characterize the entangling gates between the dressed qubits
suggest the promise of this approach both as an intermodular link
and for integration with intramodular entanglement in order to achieve
full modularity with RX qubits as well as other types of spin qubits. 

Here, we present an approach for entangling RX qubits locally within
a spin qubit module via capacitive coupling to a two-electron mediator
quantum dot driven by an ac electric field. We show that the dot-mediated
interaction is activated via the drive and generates a rapid, universal
entangling gate between the qubits. The drive serves to tailor the
spectral properties of the mediator dot such that only the two lowest-lying
zero-spin singlet states participate in mediating the qubit-qubit
interaction, while the triplet states are decoupled, potentially limiting
leakage and residual entanglement with the mediating system \cite{Srinivasa2015}.
This restructuring of the mediator dot states and energies in the
frame rotating with the drive frequency \cite{Timoney2011} serves
to simplify the effective qubit-qubit interaction by energetically
selecting specific terms, generating a single-pulse entangling gate.
By virtue of the underlying capacitive coupling mechanism and the
drive-enabled effective two-level singlet description of the mediator
dot that we describe in this work, the entangling gate does not require
the extensive sequence of pulses typically needed to mitigate leakage
in exchange-based implementations of universal two-qubit gates between
conventional exchange-only spin qubits. 

The general approach can also be applied to other spin qubits with
spin-dependent charge states that enable capacitive spin-spin coupling,
such as two-electron singlet-triplet qubits \cite{Taylor2005,Stepanenko2007,vanWeperen2011,Shulman2012,Srinivasa2015PRB,Nichol2017,Ungerer2024},
quantum dot hybrid qubits \cite{Shi2012,Mehl2015arxiv,Frees2019},
flopping-mode electron spin qubits \cite{Croot2020,Estakhri2024,Stastny2025},
flip-flop qubits \cite{Tosi2017}, and hole spin qubits \cite{Mutter2021,Yu2023,Noirot2025arxiv}.
These qubit encodings include those in which capacitive coupling requires
an additional spin-charge mixing mechanism, such as spin-orbit coupling
or magnetic gradients, that can induce mixing between the singlet
and triplet states of the mediator dot. Leakage due to this mixing
can be energetically suppressed via the effective singlet-triplet
gap enabled by the drive. We focus on RX qubits in this work as they
are exchange-only qubits that allow for universal electrical control,
and furthermore as the intrinsic spin-charge mixing present in the
logical RX qubit states enables direct capacitive coupling to the
mediator dot without inducing singlet-triplet mixing. Entanglement
mediated by an ac-driven coupler plays a significant role in two-qubit
gate approaches for superconducting qubits \cite{McKay2016,Yan2018,Krantz2019,Blais2021,Zhang2024},
and a hybrid approach has also recently been developed for using a
driven superconducting offset-charge transmon qubit as a coupler to
mediate entanglement between RX qubits via a longitudinal capacitive
interaction \cite{Kang2025}. 

In the context of achieving full modularity, the intramodular entangling
approach that we present in this work is compatible with the long-range,
sideband-based intermodular entangling gate framework of Ref. \cite{Srinivasa2024}.
We find that these two approaches give rise to the same types of gates,
as is expected from the common capacitive basis of the interactions.
In particular, since (1) both intermodular and intramodular entangling
gates are activated via driving fields, and (2) long-range entanglement
via a cavity does not require resonance with the cavity photon frequency,
distinct local and long-range coupling modes can be defined and tuned
via external drives. Furthermore, switching between the intramodular
and intermodular entangling regimes is possible by switching the corresponding
drives on or off. Thus, the approach we describe here potentially
enables the integration of entanglement across a wide range of distances
required for fully modular spin-based quantum information processing. 

\begin{figure}
\begin{centering}
\includegraphics[width=0.9\columnwidth]{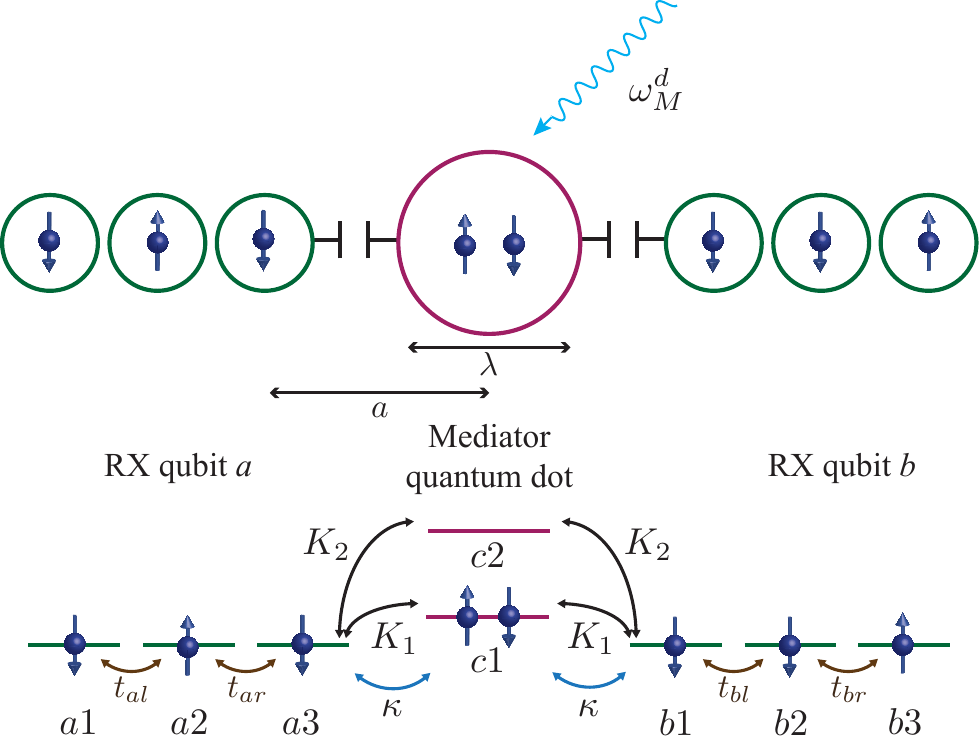}
\par\end{centering}
\caption{\label{fig:RX-MEQD}Schematic illustration of the quantum dot system
investigated in this work for driven dot-mediated capacitive interaction
of two resonant exchange (RX) qubits, along with the level diagram
and associated parameters in the extended multiorbital Hubbard model
description of the system {[}Eq. \ref{eq:Hl}{]}. The curved arrows
between orbitals in the level diagram indicate the coupling terms
in the Hamiltonian. }
\end{figure}

\section{\label{sec:Theoretical-framework}Driven dot-mediated entanglement
of resonant exchange qubits }

\subsection{\label{subsec:generalmodel}Intramodular coupling model}

To describe interactions within a spin qubit module, we consider a
pair of RX qubits \cite{Medford2013,Taylor2013,Doherty2013,Pal2014,Pal2015,Wardrop2016,Feng2021},
each encoded in three-electron spin states within a triple quantum
dot and capacitively coupled to a two-level, two-electron mediator
quantum dot \cite{Srinivasa2015} in a linear geometry (Fig. \ref{fig:RX-MEQD}).
We assume quantum dot confinement potentials such that the size $\lambda$
of the mediator dot is large compared to the sizes of the dots within
each RX qubit. Accordingly, we consider only the lowest-energy orbital
level for each dot within the qubits and the two lowest-energy orbital
levels in the center mediator dot. We also assume that the interdot
tunnel barriers within and between the two RX qubits are set such
that tunneling occurs within each RX qubit but is suppressed between
the dots of each qubit and the mediator dot \cite{Beil2014thesis}.
We write the Hamiltonian of the system as
\begin{equation}
H_{l}=H_{Q}+H_{M}+H_{QM},\label{eq:Hl}
\end{equation}
where the Hubbard model terms describing the three-electron triple
quantum dot for each RX qubit $\alpha=a,b$ are given by \cite{Taylor2013}
\begin{align}
H_{Q} & =\sum_{\alpha=a,b}H_{\alpha}\nonumber \\
 & =\sum_{\alpha=a,b}\left(H_{\alpha n}+H_{\alpha t}\right)\label{eq:HQ}
\end{align}
with 
\begin{align}
H_{\alpha n} & \equiv\sum_{i=1}^{3}\left[-\epsilon_{\alpha i}n_{\alpha i}+\frac{U_{\alpha}}{2}n_{\alpha i}\left(n_{\alpha i}-1\right)\right]\nonumber \\
 & \ \ \ \ \ \ +V_{\alpha}\left(n_{\alpha1}n_{\alpha2}+n_{\alpha2}n_{\alpha3}\right),\label{eq:Halphan}\\
H_{\alpha t} & =-\sum_{\sigma=\uparrow,\downarrow}\left(\frac{t_{\alpha l}}{\sqrt{2}}c_{\alpha2\sigma}^{\dagger}c_{\alpha1\sigma}+\frac{t_{\alpha r}}{\sqrt{2}}c_{\alpha3\sigma}^{\dagger}c_{\alpha2\sigma}+\text{{\rm H.c.}}\right).\label{eq:Halphat}
\end{align}
In Eqs. (\ref{eq:Halphan}) and (\ref{eq:Halphat}), $n_{\alpha i}=\sum_{\sigma}n_{\alpha i\sigma}=\sum_{\sigma}c_{\alpha i\sigma}^{\dagger}c_{\alpha i\sigma}$
denotes the electron number operator for dot $i,$ where $c_{\alpha i\sigma}^{\dagger}$
is the creation operator for an electron in the lowest-energy orbital
of dot $i$ with spin $\sigma=\uparrow,\downarrow,$ $-\epsilon_{\alpha i}$
is the energy of the lowest orbital level for dot $i$ that is set
via gate voltages applied to the dot, $U_{\alpha}$ and $V_{\alpha}$
are the on-site and nearest-neighbor Coulomb repulsion energies, respectively,
and $t_{\alpha l}$ ($t_{\alpha r}$) is the amplitude for tunneling
between the left (right) and center dots of RX qubit $\alpha.$ For
convenience, we use $\alpha i$ in this work to refer to the specific
dot within RX qubit $\alpha$ having the lowest-energy orbital level
$-\epsilon_{\alpha i}.$ Note that $H_{\alpha n}$ is diagonal with
respect to the charge occupation defined by the set of electron number
operator eigenvalues $\left(n_{\alpha1},n_{\alpha2},n_{\alpha3}\right)$,
and that we have chosen the signs of the orbital energies to account
for applied plunger gate voltages $P_{\alpha i}$ such that $\epsilon_{\alpha i}=eP_{\alpha i}$
and more positive gate voltages lead to lower orbital energies. In
$H_{\alpha t},$ we have defined $t_{\alpha l}$ and $t_{\alpha r}$
to be real and to represent singlet-singlet tunneling amplitudes \cite{Taylor2013}. 

\begin{figure}
\begin{centering}
\includegraphics[width=0.95\columnwidth]{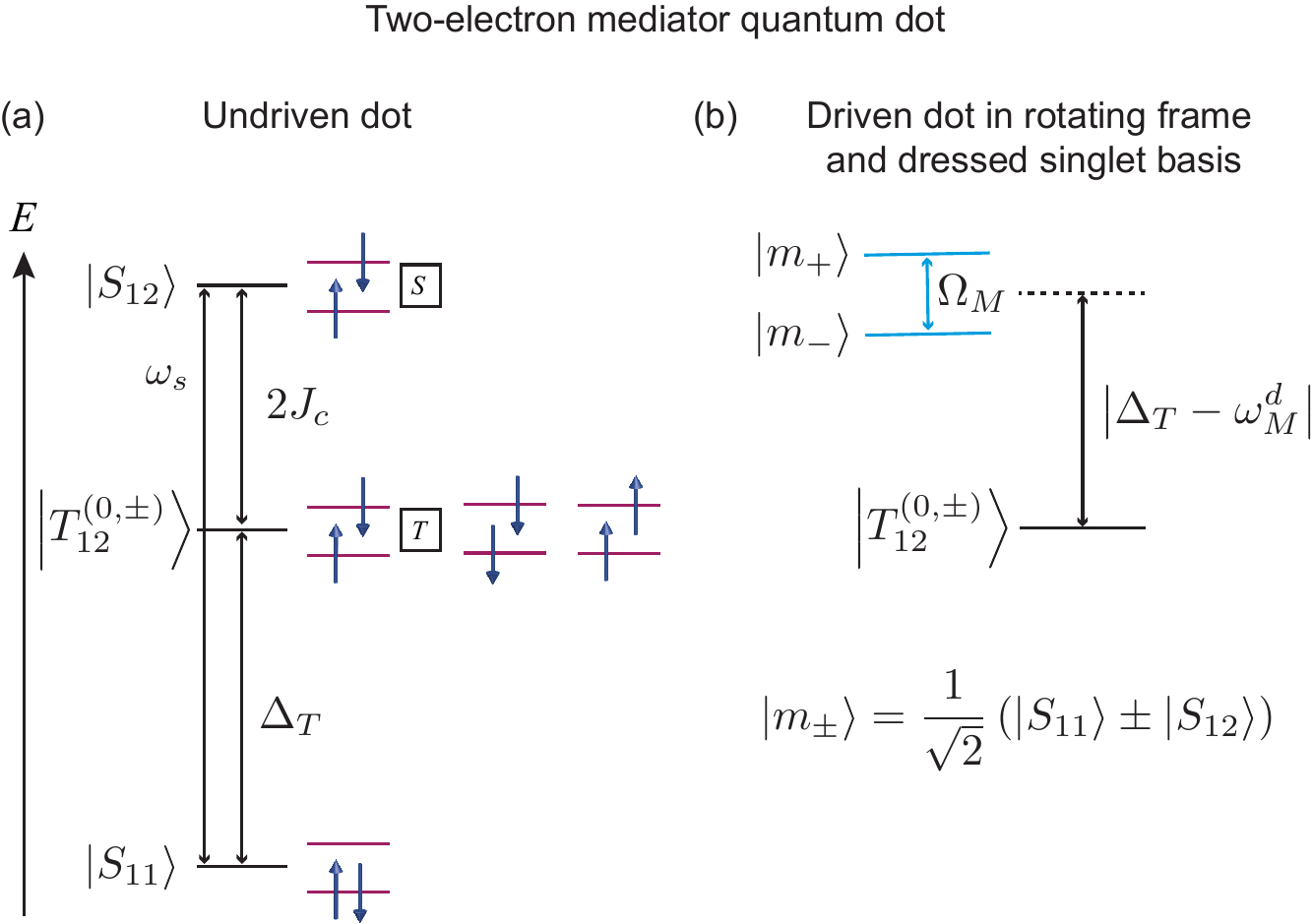}
\par\end{centering}
\caption{\label{fig:mediatordotspectrum}Low-energy spectrum of the two-level,
two-electron dot (see Appendix \ref{sec:MEQDdipole} for details)
that mediates qubit-qubit entanglement in the approach presented in
this work. (a) Spectrum of $H_{c}^{\prime}$ {[}Eq. (\ref{eq:Hcpr}){]}
in the absence of driving (adapted from Ref. \cite{Srinivasa2015}).
(b) Spectrum of the driven mediator dot in the rotating frame and
dressed singlet basis, as described by $\tilde{H}_{M}$ {[}Eq. (\ref{eq:HMtilde}){]}.}
\end{figure}

The Hamiltonian of the center two-level mediator dot in the multielectron
regime can be written in terms of a multiorbital Hubbard model as
\cite{Srinivasa2015}
\begin{equation}
H_{M}=H_{c}+H_{M}^{d}\label{eq:HM}
\end{equation}
where 
\begin{align}
H_{c} & \equiv-\sum_{j=1,2}\epsilon_{cj}n_{cj}+\frac{U_{c}}{2}n_{c1}\left(n_{c1}-1\right)\nonumber \\
 & \ \ \ \ \ \ +K_{c}n_{c1}n_{c2}+J_{c}\sum_{\sigma,\sigma'}c_{c1\sigma}^{\dag}c_{c2\sigma'}^{\dag}c_{c1\sigma'}c_{c2\sigma}\label{eq:Hc}
\end{align}
describes the mediator dot in the absence of the drive, with $n_{cj}$
and $-\epsilon_{cj}$ denoting the number operator and corresponding
energy, respectively, for orbital $cj$ and $K_{c}$ ($J_{c}$) denoting
the Coulomb repulsion (exchange) energy between two electrons occupying
different orbitals $c1$ and $c2.$ In Eq. (\ref{eq:Hc}), we have
defined $U_{c}\equiv U_{c1}$ as the on-site Coulomb repulsion energy
for two electrons in the lower level $c1$ of the mediator dot and
have set $U_{c2}=0$ to neglect double occupation of the upper mediator
dot level $c2$ \cite{Srinivasa2015}. As discussed in Appendix \ref{sec:MEQDdipole},
we have also neglected a term $H_{u}$ {[}Eq. (\ref{eq:Hu}){]} describing
on-site occupation-modulated hopping of electrons between orbitals
$c1$ and $c2$ \cite{Hubbard1963,Yang2011Hubbard,Yang2011geometry}.
The low-energy spectrum of the mediator dot in the two-electron regime
is shown in Fig. \ref{fig:mediatordotspectrum}(a). 

The second term in Eq. (\ref{eq:HM}) represents the dipole interaction
of the two-electron mediator dot with a classical external electric
field. Expressing the electric dipole operator for the mediator dot
in terms of the electron occupation of the two dot levels and working
in the low-energy two-electron subspace as described in Appendix \ref{sec:MEQDdipole},
we can write this driving term as
\begin{align}
H_{M}^{d} & =\Omega_{M}\cos\left(\omega_{M}^{d}t+\phi_{M}\right)\sum_{\sigma}\left(c_{c1\sigma}^{\dagger}c_{c2\sigma}+\text{{\rm H.c.}}\right)\nonumber \\
 & =\Omega_{M}\cos\left(\omega_{M}^{d}t+\phi_{M}\right)\left(\ket{S_{12}}\bra{S_{11}}+\ket{S_{11}}\bra{S_{12}}\right)\label{eq:HMd}
\end{align}
where $\Omega_{M}\equiv e\lambda\mathcal{E}_{M}$ is the Rabi frequency
for a mediator dot of size $\lambda,$ $\mathcal{E}_{M},$ $\omega_{M}^{d},$
and $\phi_{M}$ denote the amplitude, frequency, and phase of the
ac electric field drive, respectively, and $\ket{S_{11}}$ and $\ket{S_{12}}$
are the lowest-energy doubly and singly occupied two-electron singlet
states of the dot {[}see Fig. \ref{fig:mediatordotspectrum}(a){]}.
As we show in Appendix \ref{sec:MEQDdipole}, the drive in Eq. (\ref{eq:HMd})
also serves to simplify the description of the mediator dot to an
effective two-level system involving only the lowest-lying two-electron
singlet states {[}Fig. (\ref{fig:mediatordotspectrum})(b){]}, while
decoupling the low-lying triplet states in the absence of additional
mechanisms that that do not conserve spin. This drive-enabled reduced
description is central to the coupling mechanism we present in this
work. 

The interaction between the two RX qubits and the mediator dot is
described by the final term in Eq. (\ref{eq:Hl}). In general, this
interaction may include both tunneling and capacitive coupling terms.
Tunneling terms lead to leakage out of the two-qubit subspace via
coupling to both the singlet and triplet states of the two-electron
dot \cite{Srinivasa2015}, as is required for the conservation of
the total spin of the eight-electron system. Accordingly, we also
find that an analysis similar to Ref. \cite{Srinivasa2015} for the
system in Fig. \ref{fig:RX-MEQD} with tunneling turned on between
the qubits and the driven mediator dot in the singlet subspace yields
identical energy shifts for the logical qubit singlet and triplet
subspaces to all orders, which does not lead to an effective exchange
interaction. In order to limit leakage, we focus on capacitive coupling
in the present work. Assuming that all quantum dots are arranged in
a linear geometry (see Fig. \ref{fig:RX-MEQD}) and including the
dominant terms for capacitive coupling to the mediator dot, we can
then write 
\begin{align}
H_{QM} & =K_{1}\left(n_{a3}n_{c1}+n_{b1}n_{c1}\right)+K_{2}\left(n_{a3}n_{c2}+n_{b1}n_{c2}\right)\nonumber \\
 & -\kappa\left(n_{a3}+n_{b1}\right)\sum_{\sigma}\left(c_{c1\sigma}^{\dagger}c_{c2\sigma}+\text{{\rm H.c.}}\right)\label{eq:HQM}
\end{align}
where $K_{1}$ and $K_{2}$ denote the strengths of capacitive coupling
between electrons occupying the outer dots (orbitals $a3$ and $b1$)
of the RX qubits and electrons in levels $c1$ and $c2$ of the mediator
dot, respectively. Here, we consider an extended capacitive interaction
model relative to Ref. \cite{Srinivasa2015PRB} in which the last
term describes occupation-modulated electron hopping between the orbitals
of the mediator dot \cite{Hubbard1963,Yang2011Hubbard,Yang2011geometry}
for the dominant contributions arising from the occupation of nearest-neighbor
qubit dot levels $a3$ and $b1$ (in the absence of the on-site term
$H_{u}$ {[}Eq. (\ref{eq:Hu}){]}). Note that we assume symmetric
Coulomb interaction strengths for the capacitive coupling of the mediator
dot electrons to both RX qubits for simplicity \cite{Srinivasa2015,Malinowski2019},
as indicated in Fig. \ref{fig:RX-MEQD}.

\begin{figure*}
\begin{centering}
\includegraphics[width=0.6\paperwidth]{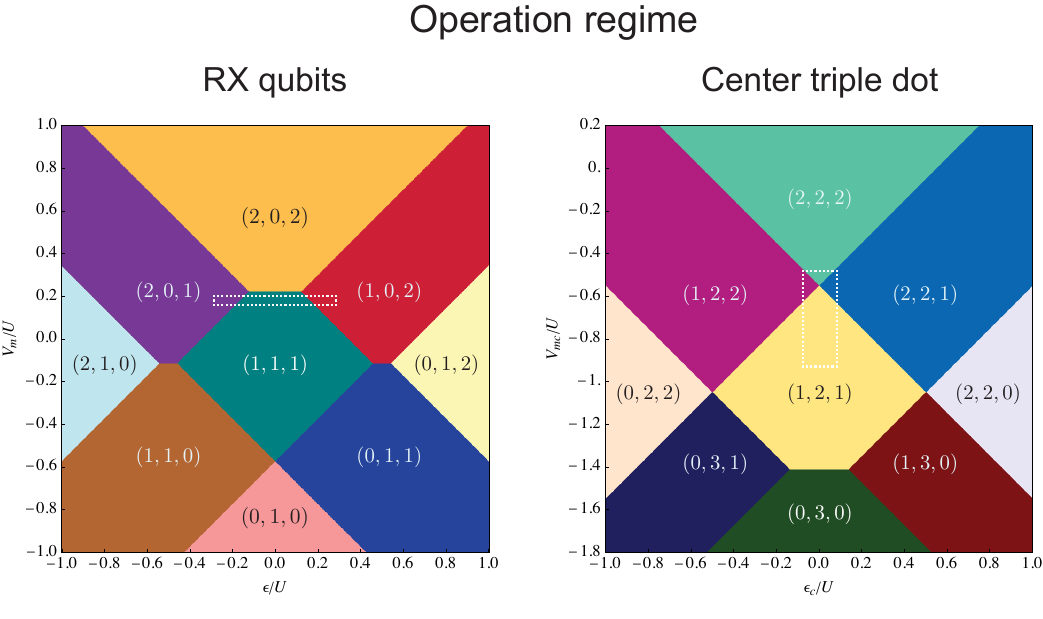}
\par\end{centering}
\caption{\label{fig:chargingdiags}Charge stability diagrams for the triple
dot systems (see Fig. \ref{fig:RX-MEQD}) associated with each RX
qubit (left), represented by the charge configuration $\left(n_{\alpha1},n_{\alpha2},n_{\alpha3}\right)$
with the axes defined by Eq. (\ref{eq:RXepsVm}), and the center three
dots (right), represented by $\left(n_{a3},n_{c},n_{b1}\right)$ with
the axes defined by Eq. (\ref{eq:centerTQDepsVm}). The RX qubit index
$\alpha$ has been suppressed in the axis labels of the left plot
for notational simplicity. The fixed parameters used in the RX qubit
triple dot (left) charging diagram are $V=0.33U$ and $\epsilon_{2}=0.90U,$
while those used in the center triple dot (right) charging diagram
are $U_{c}=0.91U,$ $V_{c}=0.28U,$ and $\epsilon_{m}=2.1U,$ where
$U\sim1\ {\rm meV}$ \cite{Malinowski2019}.}
\end{figure*}

\subsection{Charge stability diagrams and operation regime}

We next describe the regime of operation for the driven dot-mediated
entangling gate. The charge configuration for the full system in Fig.
\ref{fig:RX-MEQD} can be written in terms of the set of eigenvalues
of the number operators for electron occupation of the orbital states
within the quantum dot array, with corresponding energies specified
by Eqs. (\ref{eq:Halphan}) and (\ref{eq:Hc}). We take each triple
dot $\alpha=a,b$ to operate in the three-electron RX regime, such
that the lowest-energy configurations $\left(n_{\alpha1},n_{\alpha2},n_{\alpha3}\right)$
are $\left(1,1,1\right),$ $\left(2,0,1\right),$ and $\left(1,0,2\right).$
In this regime, the admixture of the polarized $(2,0,1)$ and $\left(1,0,2\right)$
states in the logical qubit basis states (see Appendix \ref{sec:RXqubitbas})
enables direct capacitive coupling of each RX qubit to the center
mediator dot \cite{Taylor2013,Pal2014,Pal2015,Srinivasa2016,Feng2021}.
For the two-level mediator dot, we consider the low-energy subspace
of doubly occupied and singly occupied two-electron states as shown
in Fig. \ref{fig:mediatordotspectrum}(a) and described in Appendix
\ref{sec:MEQDdipole}. We denote the charge configuration for the
full quantum dot array in the eight-electron regime by $\left(n_{a1}n_{a2}n_{a3},n_{c},n_{b1}n_{b2}n_{b3}\right),$
where $n_{c}$ represents the charge configuration in the mediator
dot. Following the notation used in Ref. \cite{Srinivasa2015}, we
use $n_{c}=2$ to denote the ground-state doubly occupied configuration
with both electrons in orbital $c1$ and $n_{c}=2^{\ast}$ to denote
the excited-state singly occupied configuration with one electron
in orbital $c1$ and one electron in orbital $c2.$ 

For the dot-mediated capacitive interaction, we consider a regime
where the Coulomb energies within the dot array are such that the
ground-state charge configuration of each RX qubit is $\left(1,1,1\right),$
while that of the mediator dot is the doubly occupied ground-state
configuration $n_{c}=2.$ Taken together with the assumption noted
in Sec. \ref{subsec:generalmodel} that tunneling occurs only within
each RX qubit, we therefore work in the subspace where the lowest-energy
configuration is $\left(111,2,111\right)$ and the charge states nearest
in energy are those with $n_{\alpha1}n_{\alpha2}n_{\alpha3}=111,201,102$
for $\alpha=a,b$ and $n_{c}=2,2^{\ast}$ for the mediator dot. 

We can visualize this charge operation regime by regarding the system
as three overlapping triple-dot systems, corresponding to $\left(n_{\alpha1},n_{\alpha2},n_{\alpha3}\right)$
for each RX qubit $\alpha$ and a center triple-dot configuration
$\left(n_{a3},n_{c},n_{b1}\right)$, and plotting charge stability
diagrams \cite{Taylor2013,Srinivasa2015} derived from the Hubbard
model terms involving occupation number operators in Eqs. (\ref{eq:Halphan}),
(\ref{eq:Hc}), and (\ref{eq:HQM}). Note that the interaction Hamiltonian
$H_{QM}$ depends only on the configuration $\left(n_{a3},n_{c},n_{b1}\right)$
of the center triple dot. In order to calculate the charge stability
diagrams, we neglect the multiorbital structure of the center dot
for simplicity and set $n_{c}=n_{c1}+n_{c2}$ such that $n_{c}$ is
the total electron occupation of the mediator dot. 

We describe the triple-dot configuration for each RX qubit via the
Hamiltonian $H_{\alpha n}$ {[}Eq. (\ref{eq:Halphan}){]} and write
a corresponding Hamiltonian for the center triple-dot configuration
as 
\begin{align}
H_{abc,n} & \equiv-\epsilon_{a3}n_{a3}-\epsilon_{b1}n_{b1}-\epsilon_{m}n_{c}\nonumber \\
 & +\frac{U_{a}}{2}n_{a3}\left(n_{a3}-1\right)+\frac{U_{b}}{2}n_{b1}\left(n_{b1}-1\right)\nonumber \\
 & +\frac{U_{c}}{2}n_{c}\left(n_{c}-1\right)\nonumber \\
 & +V_{c}\left(n_{a3}n_{c}+n_{c}n_{b1}\right),\label{eq:Habcn}
\end{align}
where we have set $U_{c2}=0$ for consistency with Eq. (\ref{eq:Hc}),
taken $K_{c}=U_{c}$ \cite{Malinowski2019}, and approximated both
$K_{1}$ and $K_{2}$ by $V_{c}.$ Here, we have also neglected the
last term in Eq. (\ref{eq:HQM}) for the calculation of the charge
stability diagrams, since typically $\kappa\ll K_{1,2}$ \cite{Hubbard1963,Yang2011Hubbard}.
We plot the resulting charge stability diagrams for both the RX qubit
and center triple dot configurations in Fig. \ref{fig:chargingdiags}
as a function of the detunings 
\begin{align}
\epsilon_{\alpha} & \equiv-\frac{1}{2}\left(\epsilon_{\alpha1}-\epsilon_{\alpha3}\right),\nonumber \\
V_{m\alpha} & \equiv-\epsilon_{\alpha2}+\frac{1}{2}\left(\epsilon_{\alpha1}+\epsilon_{\alpha3}\right)\label{eq:RXepsVm}
\end{align}
for each RX qubit $\alpha=a,b$ (see Appendix \ref{sec:RXqubitbas})
and 
\begin{align}
\epsilon_{c} & \equiv\frac{1}{2}\left(\epsilon_{a3}-\epsilon_{b1}\right),\nonumber \\
V_{mc} & \equiv-\epsilon_{m}+\frac{1}{2}\left(\epsilon_{a3}+a_{b1}\right)\label{eq:centerTQDepsVm}
\end{align}
for the center triple dot. 

The operation regime for the full system is defined by identifying
simultaneous operation points in both diagrams. These operation points
are related via the constraint 
\begin{equation}
V_{mc}+\epsilon_{m}=V_{m\alpha}+\epsilon_{\alpha2},\label{eq:chargingdiagconstr}
\end{equation}
which is derived based on the common orbital levels of dots $a3$
and $b1$ in the overlapping triple-dot systems and assumes for simplicity
that the parameters for the two RX qubits are identical, such that
$U_{a}=U_{b}\equiv U,$ $V_{a}=V_{b}\equiv V,$ $\epsilon_{a2}=\epsilon_{b2}\equiv\epsilon_{2},$
$\epsilon_{a}=\epsilon_{b}\equiv\epsilon,$ and $\Delta_{a}=\Delta_{b}\equiv\Delta.$
For this case, we find that $\epsilon_{c}=\epsilon.$ An example of
simultaneous operation points for the coupling approach discussed
in this work is indicated by the dotted rectangles in Fig. \ref{fig:chargingdiags},
where combining the relevant charge configurations for each RX qubit
triple dot with those for the center triple dot yields the regime
of full configurations $\left(n_{a1}n_{a2}n_{a3},n_{c},n_{b1}n_{b2}n_{b3}\right)$
described above. Quantitatively, working near $\epsilon=\epsilon_{c}=0$
and setting $\Delta=0.12U$ gives $V_{ma}=V_{mb}\equiv V_{m}=U-2V-\Delta=0.22U$
using the definition of $\Delta$ given in Appendix \ref{sec:RXqubitbas},
and thus $V_{mc}=-0.98U$ from Eq. (\ref{eq:chargingdiagconstr})
for the parameters used in Fig. \ref{fig:chargingdiags}. 

\subsection{\label{subsec:EffHamiltonian}Effective Hamiltonian model}

We now derive an effective Hamiltonian model for the intramodular
coupling described by Eq. (\ref{eq:Hl}) in the operation regime shown
in Fig. \ref{fig:chargingdiags}. This model generates the driven
dot-mediated entangling gates. Here, we consider symmetric operation
of the RX qubits (Appendix \ref{sec:RXqubitbas}) and set $\epsilon_{a}=\epsilon_{b}=0,$
where the qubits possess first-order insensitivity to charge noise,
as well as $t_{\alpha l}=t_{\alpha r}\equiv t_{\alpha}$ for $\alpha=a,b$
\cite{Srinivasa2016}. These fixed operation points are consistent
with those chosen for the qubits in the sideband-based long-range
entangling approach of Ref. \cite{Srinivasa2024}. 

In Appendix \ref{sec:RXqubitbas}, we show how the Hubbard model description
of a triple quantum dot in the three-electron regime used to define
the RX qubit {[}Eqs. (\ref{eq:HQ})-\ref{eq:Halphat}{]} can be reduced
to an effective two-level model for the chosen operation regime. Combining
Eq. (\ref{eq:HQeff}) with the effective two-level approximation for
the driven two-electron mediator dot discussed in Appendix \ref{sec:MEQDdipole}
{[}see Eq. (\ref{eq:HMeff}) and Fig. \ref{fig:mediatordotspectrum}(b){]}
yields an effective Hamiltonian for $H_{l}$ {[}Eq. (\ref{eq:Hl}){]}
given by
\begin{align}
H_{l}^{{\rm eff}} & =H_{Q}^{{\rm eff}}+H_{M}^{{\rm eff}}+H_{QM}^{{\rm eff}}\nonumber \\
 & =\sum_{\alpha=a,b}\frac{\omega_{\alpha}}{2}\sigma_{\alpha}^{z}+\frac{\omega_{s}}{2}\tau_{z}+\Omega_{M}\cos\left(\omega_{M}^{d}t+\phi_{M}\right)\tau_{x}\nonumber \\
 & +\sum_{\alpha=a,b}\left(q_{0}^{\alpha}\mathbf{1}-q_{z}^{\alpha}\sigma_{\alpha}^{z}-q_{x}^{\alpha}\sigma_{\alpha}^{x}\right)\nonumber \\
 & \times\left(K_{0}\mathbf{1}+\frac{\Delta K}{2}\tau_{z}-K_{m}\tau_{x}\right).\label{eq:Hleff}
\end{align}
In writing Eq. (\ref{eq:Hleff}), we have used Eqs. (\ref{eq:n1n3deff})
and (\ref{eq:nc1nc2}) to express $H_{QM}$ {[}Eq. (\ref{eq:HQM}){]}
in terms of the two-level approximations for the qubits and the mediator
dot and have defined $K_{0}\equiv\left(3K_{1}+K_{2}\right)/2,$ $\Delta K\equiv K_{2}-K_{1},$
and $K_{m}\equiv\sqrt{2}\kappa.$ As described in Appendix \ref{sec:RXqubitbas},
the specific dependence of $H_{QM}^{{\rm eff}}$ on RX qubit parameters
is contained in the coefficients $q_{0,z,x}^{\alpha}$ used to express
the number operators in the qubit basis. By replacing these coefficients
with the corresponding coefficients for other types of spin qubits
that can be capacitively coupled, the theory developed in this work
can be adapted to a wide variety of spin qubit systems as noted in
Sec. \ref{sec:Introduction}. For those qubits that require additional
spin-charge mixing mechanisms such as spin-orbit interaction or magnetic
gradients to achieve capacitive coupling, these mechanisms would accordingly
be incorporated into the specific forms of the qubit-dependent coefficients.
Note that in these cases, the entangling method we present here can
be implemented provided the spin-charge mixing mechanism is sufficiently
weak in the region of the mediator dot compared to $\left|\Delta_{T}-\omega_{M}^{d}\right|$
(see Fig. \ref{fig:mediatordotspectrum} and Sec. \ref{sec:MEQDdipole})
in order to avoid leakage due to induced mixing between the mediator
dot singlet and triplet states and maintain the validity of the two-level
singlet description. Here, we focus on RX qubits as they enable direct
capacitive coupling via intrinsic spin-charge mixing \cite{Taylor2013,Pal2014,Pal2015,Feng2021},
which does not cause singlet-triplet mixing in the mediator dot. 

We can rewrite Eq. (\ref{eq:Hleff}) as
\begin{align}
H_{l}^{{\rm eff}} & =\sum_{\alpha=a,b}\frac{\omega_{\alpha}^{\prime}}{2}\sigma_{\alpha}^{z}+\frac{\omega_{s}^{\prime}}{2}\tau_{z}+\Omega_{M}\cos\left(\omega_{M}^{d}t\right)\tau_{x}\nonumber \\
 & -\sum_{\alpha=a,b}q_{0}^{\alpha}K_{m}\tau_{x}\nonumber \\
 & -\sum_{\alpha=a,b}q_{z}^{\alpha}\sigma_{\alpha}^{z}\left(\frac{\Delta K}{2}\tau_{z}-K_{m}\tau_{x}\right)\nonumber \\
 & -\sum_{\alpha=a,b}q_{x}^{\alpha}\sigma_{\alpha}^{x}\left(K_{0}\mathbf{1}+\frac{\Delta K}{2}\tau_{z}-K_{m}\tau_{x}\right),\label{eq:Hleffshiftfreq}
\end{align}
where we have set $\phi_{M}=0$ for simplicity, defined the modified
qubit and mediator dot frequencies
\begin{align}
\omega_{\alpha}^{\prime} & =\omega_{\alpha}-2q_{z}^{\alpha}K_{0},\nonumber \\
\omega_{s}^{\prime} & =\omega_{s}+\sum_{\alpha=a,b}q_{0}^{\alpha}\Delta K,\label{eq:omegaprfreqs}
\end{align}
and dropped terms proportional to the identity operator. Note that
both frequency shifts in Eq. (\ref{eq:omegaprfreqs}) arise from the
capacitive interaction between the qubits and the mediator dot {[}Eq.
(\ref{eq:HQM}){]}. 

To identify terms relevant for the low-energy dynamics of the system,
we now transform Eq. (\ref{eq:Hleffshiftfreq}) to a frame rotating
at the mediator dot drive frequency $\omega_{M}^{d}$ via the unitary
transformation 
\begin{equation}
U_{{\rm rf}}=e^{-i\omega_{M}^{d}t\left(\sum_{\alpha}\sigma_{\alpha}^{z}+\tau_{z}\right)/2}.\label{eq:Urf}
\end{equation}
Writing $H_{l,{\rm rf}}^{{\rm eff}}\equiv U_{{\rm rf}}^{\dagger}H_{l}^{{\rm eff}}U_{{\rm rf}}-iU_{{\rm rf}}^{\dagger}\dot{U}_{{\rm rf}},$
dropping rapidly oscillating terms $\sim e^{\pm i\omega_{M}^{d}t}$
and $\sim e^{\pm2i\omega_{M}^{d}t}$ for $\Omega_{M}\ll\omega_{M}^{d},$
and considering the case of a resonantly driven mediator dot for simplicity
such that $\omega_{M}^{d}=\omega_{s}^{\prime},$ we find 
\begin{align}
H_{l,{\rm rf}}^{{\rm eff}} & \approx\sum_{\alpha=a,b}\frac{\delta_{\alpha}}{2}\sigma_{\alpha}^{z}+\frac{\Omega_{M}}{2}\tau_{x}\nonumber \\
 & -\frac{\Delta K}{2}\sum_{\alpha=a,b}q_{z}^{\alpha}\sigma_{\alpha}^{z}\tau_{z}\nonumber \\
 & +K_{m}\sum_{\alpha=a,b}q_{x}^{\alpha}\left(\sigma_{\alpha}^{+}\tau_{-}+\sigma_{\alpha}^{-}\tau_{+}\right)\label{eq:Hleffrf}
\end{align}
where we have defined the qubit-drive frequency detuning $\delta_{\alpha}\equiv\omega_{\alpha}^{\prime}-\omega_{M}^{d}.$
We next rotate the driven mediator dot to a dressed singlet representation
via 
\begin{equation}
U_{s}=e^{-i\pi\tau_{y}/4},\label{eq:Us}
\end{equation}
such that the Hamiltonian becomes
\begin{align}
H_{l,{\rm s}}^{{\rm eff}} & =H_{0}+V,\nonumber \\
H_{0} & \equiv\sum_{\alpha=a,b}\frac{\delta_{\alpha}}{2}\sigma_{\alpha}^{z}+\frac{\Omega_{M}}{2}\tilde{\tau}_{z},\nonumber \\
V & \equiv\frac{\Delta K}{2}\sum_{\alpha=a,b}q_{z}^{\alpha}\sigma_{\alpha}^{z}\tilde{\tau}_{x}\nonumber \\
 & +\frac{K_{m}}{2}\sum_{\alpha=a,b}q_{x}^{\alpha}\left[\sigma_{\alpha}^{+}\left(\tilde{\tau}_{z}-i\tilde{\tau}_{y}\right)+{\rm H.c.}\right],\label{eq:Hleffs}
\end{align}
where $\tilde{\tau}_{z}\equiv\ket{m_{+}}\bra{m_{+}}-\ket{m_{-}}\bra{m_{-}}$
with $\ket{m_{\pm}}\equiv\left(\ket{S_{11}}\pm\ket{S_{12}}\right)/\sqrt{2}$
denoting the dressed singlet states {[}see Fig. \ref{fig:mediatordotspectrum}(b){]}.
Using dressed states for the mediator dot enables suppression of charge
dephasing \cite{Timoney2011,Laucht2017} between the singlet states
$\ket{S_{11}}$ and $\ket{S_{12}}.$ Since oscillations between these
two states occur with the Rabi frequency $\Omega_{M}$ {[}see Eq.
(\ref{eq:HMeff}){]} and dephasing in this original singlet basis
translates to a transition between the dressed singlet states $\ket{m_{\pm}}$
of the mediator dot, dephasing in the original basis can be suppressed
for the dressed singlet states as long as the dephasing rate $\gamma_{M}\ll\Omega_{M}.$
This condition is satisfied for the Rabi frequency $\Omega_{M}$ determined
in Sec. \ref{subsec:parameters_Kab} and the dephasing rates $\gamma_{M}$
used to calculate the fidelity in Fig. \ref{fig:Fvsdephrates}.

In an interaction picture with respect to $H_{0}$ obtained by applying
$U_{{\rm 0}}=e^{-iH_{0}t},$ we find terms in the Hamiltonian with
time-dependent factors $e^{\pm i\Omega_{M}t},e^{\pm i\delta_{\alpha}t},e^{\pm i\left(\delta_{\alpha}+\Omega_{M}\right)t},$
and $e^{\pm i\left(\delta_{\alpha}-\Omega_{M}\right)t}.$ Applying
a rotating wave approximation for $\Omega_{M}\ll\left|\delta_{\alpha}\right|,\left|\delta_{\alpha}\pm\Omega_{M}\right|,$
which is satisfied for typical system parameters as described in Appendix
\ref{sec:RWAvalidity}, we can neglect rapidly oscillating terms and
transform out of the interaction picture to obtain
\begin{equation}
H_{l,{\rm RWA}}^{{\rm eff}}=\sum_{\alpha=a,b}\frac{\delta_{\alpha}}{2}\sigma_{\alpha}^{z}+\frac{\Omega_{M}}{2}\tilde{\tau}_{z}+\frac{\Delta K}{2}\sum_{\alpha=a,b}q_{z}^{\alpha}\sigma_{\alpha}^{z}\tilde{\tau}_{x}.\label{eq:HleffRWA}
\end{equation}
We see from Eq. (\ref{eq:HleffRWA}) that, in addition to reducing
the two-electron mediator dot description to an effective two-level
system as described in Appendix \ref{sec:MEQDdipole}, driving the
mediator dot also simplifies the form of the interaction with the
qubits via a separation of energy scales for $\Omega_{M}\ll\left|\delta_{\alpha}\right|,\left|\delta_{\alpha}\pm\Omega_{M}\right|$
(see Fig. \ref{fig:mediatordotspectrum} and Sec. \ref{sec:RWAvalidity}). 

The matrix representation of the Hamiltonian in Eq. (\ref{eq:HleffRWA})
is block-diagonal, with four two-dimensional blocks in the low-energy
mediator dot subspace spanned by $\left\{ \ket{m_{+}},\ket{m_{-}}\right\} $
that are each associated with one of the two-qubit states $\ket{00},\ket{01},\ket{10},$
or $\ket{11}.$ We diagonalize this Hamiltonian via a qubit state-conditional
rotation of the mediator dot subspace, given by
\begin{align}
U_{d} & =e^{-i\left[\left(\theta_{s}+\theta_{d}\right)\sigma_{a}^{z}+\left(\theta_{s}-\theta_{d}\right)\sigma_{b}^{z}\right]\tilde{\tau}_{y}/4}\label{eq:Ud}
\end{align}
where $\tan\theta_{s}\equiv\tan\theta_{11}=-\tan\theta_{00}=\left(q_{z}^{a}+q_{z}^{b}\right)\Delta K/\Omega_{M}$
and $\tan\theta_{d}\equiv\tan\theta_{10}=-\tan\theta_{01}=\left(q_{z}^{a}-q_{z}^{b}\right)\Delta K/\Omega_{M}$
define the angles of rotation for each two-qubit state. Noting that
$\left[U_{d},\sigma_{\alpha}^{z}\right]=0$ for $\alpha=a,b,$ this
transformation yields 
\begin{align}
H_{l,d}^{{\rm eff}} & \equiv U_{d}^{\dagger}H_{l,{\rm RWA}}^{{\rm eff}}U_{d}\nonumber \\
 & =\sum_{\alpha=a,b}\frac{\delta_{\alpha}}{2}\sigma_{\alpha}^{z}+\frac{\Omega_{M}^{\prime}}{2}\tau_{z}^{M}+\mathcal{K}_{ab}\sigma_{a}^{z}\sigma_{b}^{z}\tau_{z}^{M}\nonumber \\
 & =\sum_{\alpha=a,b}\frac{\delta_{\alpha}}{2}\sigma_{\alpha}^{z}+\frac{\hat{\Omega}_{M}}{2}\tau_{z}^{M},\label{eq:Hleffd}
\end{align}
where $\hat{\Omega}_{M}\equiv\Omega_{M}^{\prime}+2\mathcal{K}_{ab}\sigma_{a}^{z}\sigma_{b}^{z}$
is an operator that describes a qubit state-dependent frequency for
the mediator dot, with $\Omega_{M}^{\prime}\equiv\left(\Omega_{s}+\Omega_{d}\right)/2,$
$\mathcal{K}_{ab}\equiv\left(\Omega_{s}-\Omega_{d}\right)/4,$ and
$\Omega_{s\left(d\right)}\equiv\sqrt{\Omega_{M}^{2}+\left(q_{z}^{a}\pm q_{z}^{b}\right)^{2}\Delta K^{2}},$
while $\tau_{z}^{M}$ is a Pauli $z$ operator describing the mediator
dot in the diagonal basis obtained via Eq. (\ref{eq:Ud}) with eigenvalues
$\pm1.$ In the low-energy subspace for the mediator dot corresponding
to the replacement $\tau_{z}^{M}\rightarrow-1,$ Eq. (\ref{eq:Hleffd})
takes the form (dropping terms proportional to the identity operator
within this subspace)
\begin{equation}
H_{l,d}^{{\rm \left(-\right)}}\equiv\sum_{\alpha=a,b}\frac{\delta_{\alpha}}{2}\sigma_{\alpha}^{z}-\mathcal{K}_{ab}\sigma_{a}^{z}\sigma_{b}^{z}\label{eq:Hldminus}
\end{equation}
and directly generates a two-qubit controlled-phase gate when $\delta_{\alpha}=2\mathcal{K}_{ab}$
for $\alpha=a,b,$ with phase $\varphi=4\mathcal{K}_{ab}t$ \cite{Vandersypen2005,Srinivasa2015PRB,Srinivasa2024}.
Note that, since the full-space Hamiltonian $H_{l,d}^{{\rm eff}}$
{[}Eq. (\ref{eq:Hleffd}){]} is already diagonal, Eq. (\ref{eq:Hldminus})
is obtained directly from $H_{l,d}^{{\rm eff}}$ without perturbation
theory. This qubit-qubit interaction is activated via the mediator
dot drive {[}Eq. (\ref{eq:HMd}){]}, which gives rise to the qubit
state-conditional rotation $U_{d},$ and does not exist in the absence
of driving. 

Finally, we transform the Hamiltonian $H_{l,d}^{{\rm eff}}$ to a
dressed-state basis $\left\{ \ket{e}_{\alpha},\ket{g}_{\alpha}\right\} $
for the qubits {[}Eq. (\ref{eq:egdrRXbas}){]} via $U_{q}$ {[}Eq.
(\ref{eq:Uq}){]} as described in Appendix \ref{sec:RXqubitbas} in
order to consider the effective dynamics generated by the local coupling
Hamiltonian $H_{l}$ in the dressed-qubit basis used in Ref. \cite{Srinivasa2024}
for long-range cavity-mediated entangling gates. We find
\begin{align}
H_{q} & \equiv U_{q}^{\dagger}H_{l,d}^{{\rm eff}}U_{q}\nonumber \\
 & =-\sum_{\alpha=a,b}\frac{\delta_{\alpha}}{2}\tilde{\sigma}_{\alpha}^{x}+\frac{\Omega_{M}^{\prime}}{2}\tau_{z}^{M}+\mathcal{K}_{ab}\tilde{\sigma}_{a}^{x}\tilde{\sigma}_{b}^{x}\tau_{z}^{M},\label{eq:Hq}
\end{align}
Note that $H_{q}$ is block-diagonal, with two decoupled four-dimensional
blocks corresponding to the eigenvalues $\pm1$ of $\tau_{z}^{M}.$
Thus, we can again derive an effective two-qubit interaction Hamiltonian
from the full-space Hamiltonian $H_{q}$ without perturbation theory
in the dressed-qubit basis. As before, considering the low-energy
subspace for the mediator dot by making the replacement $\tau_{z}^{M}\rightarrow-1$
and dropping terms proportional to the identity operator within this
subspace leads to the effective two-qubit Hamiltonian
\begin{equation}
H_{-}\equiv-\sum_{\alpha=a,b}\frac{\delta_{\alpha}}{2}\tilde{\sigma}_{\alpha}^{x}-\mathcal{K}_{ab}\tilde{\sigma}_{a}^{x}\tilde{\sigma}_{b}^{x}\label{eq:Hminus}
\end{equation}
that generates the dynamics $U_{-}\left(t\right)\equiv e^{-iH_{-}t}.$
The qubit-qubit interaction term in this Hamiltonian is known to generate
a Mølmer-Sørensen gate \cite{Sorensen1999}. This gate serves as the
most widely used universal two-qubit entangling gate in platforms
for trapped-ion quantum information processing, where the Coulomb
interaction also serves as the fundamental basis for the entanglement.
For $\mathcal{K}_{ab}t=\pi\left(8m+1\right)/4$ and $\delta_{a}=\delta_{b}=8r\mathcal{K}_{ab}$
with $m,r$ denoting integers, $U_{-}$ becomes
\begin{equation}
U_{xx}=\frac{1}{\sqrt{2}}\left(\begin{array}{cccc}
1 & 0 & 0 & i\\
0 & 1 & i & 0\\
0 & i & 1 & 0\\
i & 0 & 0 & 1
\end{array}\right).\label{eq:Uxx}
\end{equation}
In terms of the gates defined in Ref. \cite{Srinivasa2024}, $U_{xx}$
is equivalent to the gate $U_{i{\rm SW}}^{1/2}$ within the two-qubit
subspace $\left\{ \ket{eg},\ket{ge}\right\} $ (also known as a $\sqrt{i{\rm SWAP}}$
gate) and equivalent to the gate $U_{i{\rm DE}}^{1/2}$ within the
subspace $\left\{ \ket{ee},\ket{gg}\right\} $ (also known as a $\sqrt{b{\rm SWAP}}$
gate), both of which are universal entangling gates \cite{Imamoglu1999,Schuch2003,Zhang2024}.
Thus, these intramodular entangling gates are of the same type as
those generated by the cavity photon-mediated long-range entangling
interactions between driven qubits discussed in Ref. \cite{Srinivasa2024}.
We note that both the local gates discussed in this work and the long-range
cavity-mediated entangling gates are activated via driving fields,
enabling switching between the local and long-range modes of coupling
via the drives without tuning either the qubit or coupler away from
optimal operation points (see Appendix \ref{sec:RXqubitbas}) while
carrying out the entangling gates. As we discuss in Sec. \ref{sec:spinqubitmodularity},
these features are favorable for integrating the local and long-range
approaches to achieve modularity for spin qubits. 

\subsection{\label{subsec:parameters_Kab}Driven dot-mediated qubit interaction
strength}

The strength $\mathcal{K}_{ab}$ of the effective driven dot-mediated
capacitive interaction between the qubits in Eq. (\ref{eq:Hminus})
sets the rate of the generated two-qubit entangling gate $U_{xx}$
{[}Eq. (\ref{eq:Uxx}){]}. We now estimate $\mathcal{K}_{ab}$ and
its scaling with the mediator dot size $\lambda$ and qubit-mediator
interdot distance $a$ (see Fig. \ref{fig:RX-MEQD}) by using Fock-Darwin
states for the dot orbital levels, as described in Appendix \ref{sec:MEQDdipole}
for the mediator dot, along with a multipole expansion of the Coulomb
interaction matrix elements $K_{1}$ and $K_{2}$ in the capacitive
coupling Hamiltonian $H_{QM}$ {[}Eq. (\ref{eq:HQM}){]} up to quadrupole
order \cite{Gamble2012,Srinivasa2015PRB}. 

Since we assume symmetric Coulomb interactions {[}see Eq. (\ref{eq:HQM}){]},
we use dot $a3$ in the left RX qubit to calculate $K_{1}$ and $K_{2}$
and take the same results to hold for the replacement $a3\rightarrow b1.$
Writing positions in terms of the coordinates $\left(x,y\right),$
we take the the quantum dot array axis to lie along the $x$ axis
with the mediator dot centered at ${\bf R}_{c}=\left(0,0\right)$
and dot $a3$ centered at the location ${\bf R}_{a3}=\left(-a,0\right).$
The wave functions for the dots are given by
\begin{equation}
\psi_{a3}\left(x,y\right)=\frac{e^{-\left[\left(x+a\right)^{2}+y^{2}\right]/4\sigma^{2}}}{\sqrt{2\pi}\sigma}\label{eq:psia3}
\end{equation}
along with $\psi_{c1}\left(x,y\right)$ and $\psi_{c2}\left(x,y\right),$
where $\psi_{c1}$ and $\psi_{c2}$ are the mediator dot orbital wave
functions given in Eq. (\ref{eq:psic1c2}). In terms of these wave
functions, the Coulomb matrix elements $K_{i}$ for $i=1,2$ are given
by $K_{i}=e^{2}K_{i}^{\prime}/4\pi\bar{\epsilon},$ where \cite{Hubbard1963,Srinivasa2015PRB}
\begin{align}
K_{i}^{\prime} & =\bra{a3,ci}\frac{1}{\left|{\bf r}-{\bf r}^{\prime}\right|}\ket{a3,ci}\nonumber \\
 & =\int\frac{\left|\psi_{a3}\left({\bf r}\right)\right|^{2}\left|\psi_{ci}\left({\bf r}^{\prime}\right)\right|^{2}}{\left|{\bf r}-{\bf r}^{\prime}\right|}d{\bf r}d{\bf r}^{\prime}\label{eq:Kimatel}
\end{align}
and $\bar{\epsilon}$ denotes the dielectric permittivity. Here, we
consider silicon quantum dots and take $\bar{\epsilon}=\bar{\epsilon}_{{\rm Si}}=11.7\bar{\epsilon}_{0}$
\cite{Gamble2012}, where $\bar{\epsilon}_{0}$ is the vacuum permittivity.
Assuming $\lambda/2a\ll1,$ we can approximate the denominator in
Eq. (\ref{eq:Kimatel}) via a multipole expansion. As restricting
the approximation to the leading-order term gives $K_{1}=K_{2}$ and
thus $\Delta K=0,$ and since the matrix element of the dipole term
vanishes, we estimate $K_{1}$ and $K_{2}$ by keeping terms up to
quadrupole order in the expansion. We then have
\begin{align}
\frac{1}{\left|{\bf r}-{\bf r}^{\prime}\right|} & \approx\frac{1}{\left|{\bf R}_{a3}-{\bf R}_{c}-{\bf b}\right|}\nonumber \\
 & =\frac{1}{\left|{\bf R}-{\bf b}\right|}\nonumber \\
 & \approx\frac{1}{a}-\frac{x^{\prime}}{a^{2}}+\frac{1}{2a^{3}}\left(2x^{\prime2}-y^{\prime2}\right)\label{eq:1byr}
\end{align}
In writing Eq. (\ref{eq:1byr}), we have for simplicity assumed $\sigma\ll\lambda$
and neglected the spatial dependence of the electron wave function
for dot $a3.$ Accordingly, we have set ${\bf r}={\bf R}_{a3}$ and
${\bf r}^{\prime}={\bf R}_{c}+{\bf b},$ where ${\bf b}$ represents
the electron position in the mediator dot relative to the dot center
${\bf R}_{c}.$ For ${\bf R}_{c}=\left(0,0\right),$ ${\bf b}={\bf r}^{\prime}=\left(x^{\prime},y^{\prime}\right).$
We have also defined the vector ${\bf R}\equiv{\bf R}_{a3}-{\bf R}_{c}$
with magnitude $R=a\gg b\sim\lambda/2.$ 
\begin{center}
\begin{figure}
\begin{centering}
\includegraphics[width=1\columnwidth]{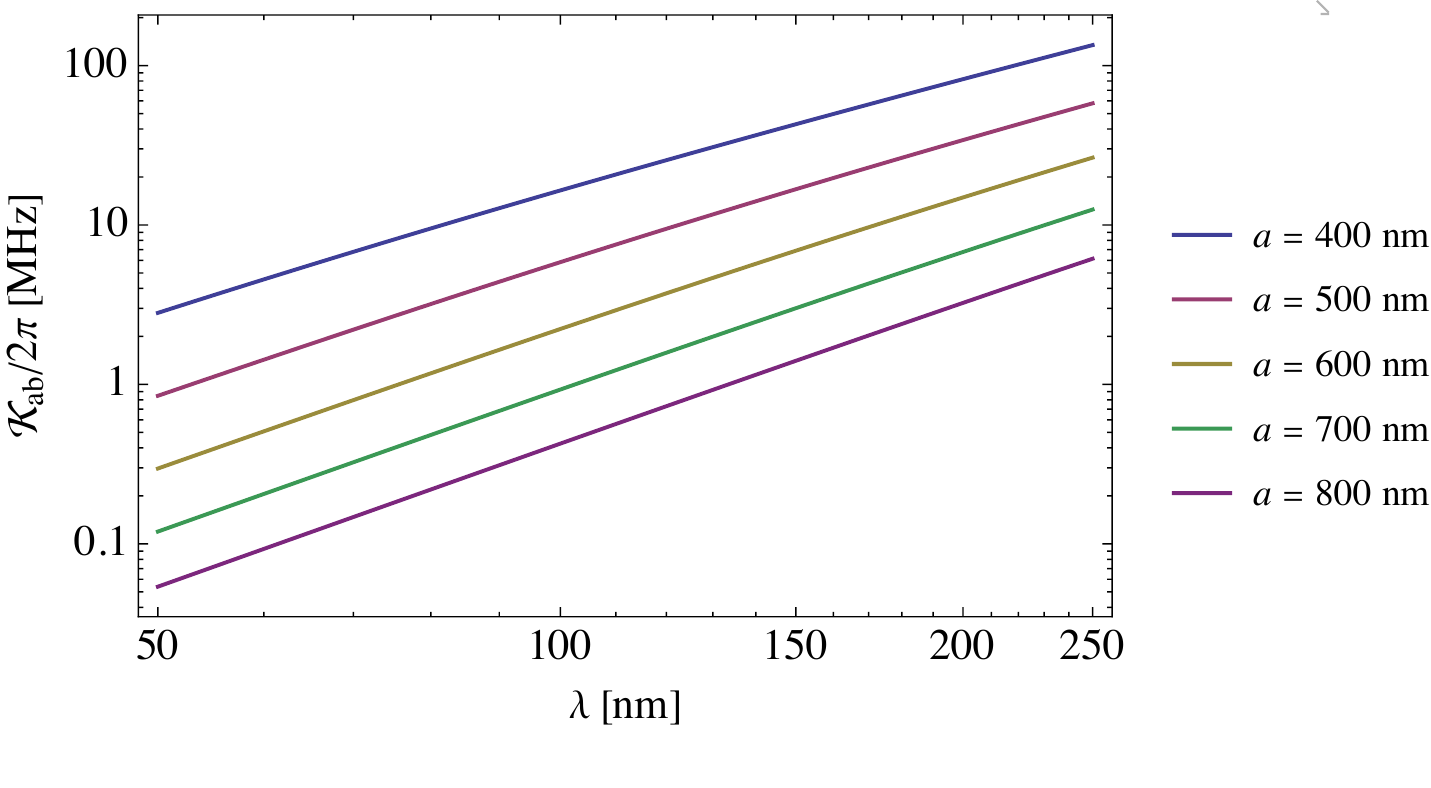}
\par\end{centering}
\caption{\label{fig:couplingstrength}Strength $\mathcal{K}_{ab}$ of the driven
dot-mediated capacitive interaction between RX qubits as a function
of mediator dot size $\lambda$ for multiple values of the qubit-mediator
dot separation $a$ and fixed electric field drive amplitude $\mathcal{E}_{M}=2\ {\rm V/m},$
calculated using the analysis and parameters given in Sec. \ref{subsec:parameters_Kab}.}
\end{figure}
\par\end{center}

Substituting Eq. (\ref{eq:1byr}) into Eq. (\ref{eq:Kimatel}) and
integrating, we find 
\begin{align}
K_{1}^{\prime} & \approx\frac{1}{a}+\frac{\lambda^{2}}{2a^{3}},\nonumber \\
K_{2}^{\prime} & \approx\frac{1}{a}+\frac{\lambda^{2}}{a^{3}},\label{eq:K1pK2papprox}
\end{align}
which we use to approximate $K_{i}=e^{2}K_{i}^{\prime}/4\pi\bar{\epsilon}_{{\rm Si}}$
for $i=1,2,$ yielding 
\begin{align}
\Delta K & \equiv K_{2}-K_{1}\nonumber \\
 & \approx\frac{e^{2}}{4\pi\bar{\epsilon}_{{\rm Si}}}\left(\frac{\lambda^{2}}{2a^{3}}\right).\label{eq:deltaK}
\end{align}
We use this expression to estimate $\mathcal{K}_{ab}.$ Note that
$\Delta K$ arises entirely from the quadrupole ($\sim1/a^{3}$) terms
in Eqs. (\ref{eq:1byr}) and (\ref{eq:K1pK2papprox}), potentially
allowing for reduced sensitivity of the mediated capacitive coupling
mechanism in this approach to charge noise relative to direct capacitive
coupling between qubits. A similar calculation for the occupation-modulated
hopping matrix element in Eq. (\ref{eq:HMd}) can be carried out using
Eq. (\ref{eq:1byr}) and gives \cite{Hubbard1963,Yang2011geometry}
$\kappa\equiv-\left(e^{2}/4\pi\bar{\epsilon}\right)\bra{a3,c1}\left|{\bf r}-{\bf r}^{\prime}\right|^{-1}\ket{a3,c2}\approx\left(e^{2}/4\pi\bar{\epsilon}\right)\left(\lambda/\sqrt{2}a^{2}\right).$
As shown in Sec. \ref{subsec:EffHamiltonian}, the associated term
in Eq. (\ref{eq:HQM}) is energetically suppressed in the effective
mediated interaction by the drive on the mediator dot and is not used
to estimate $\mathcal{K}_{ab}.$

In Sec. \ref{subsec:EffHamiltonian}, we determined the strength of
the qubit-qubit coupling in Eq. (\ref{eq:Hminus}) to be given by
$\mathcal{K}_{ab}\equiv\left(\Omega_{s}-\Omega_{d}\right)/4,$ where
$\Omega_{s\left(d\right)}\equiv\sqrt{\Omega_{M}^{2}+\left(q_{z}^{a}\pm q_{z}^{b}\right)^{2}\Delta K^{2}},$
$\Omega_{M}\equiv e\lambda\mathcal{E}_{M}$ is the Rabi frequency
for the mediator dot driving field defined in Appendix \ref{sec:MEQDdipole},
and $q_{z}^{\alpha}=q_{z}^{\alpha}\left(\Delta_{\alpha},t_{\alpha}\right)$
for $\alpha=a,b$ are the dimensionless qubit parameter-dependent
coefficients given in Eq. (\ref{eq:q0qzqx}). Note that in the limit
$\left(q_{z}^{a}\pm q_{z}^{b}\right)\Delta K\ll\Omega_{M},$ a Schrieffer-Wolff
transformation of the effective Hamiltonian in Eq. (\ref{eq:HleffRWA})
can be performed and leads to an expression of the same form as Eq.
(\ref{eq:Hminus}) with $\mathcal{K}_{ab}\approx q_{z}^{a}q_{z}^{b}\Delta K^{2}/2\Omega_{M}\sim\lambda^{3}/a^{6}\mathcal{E}_{M}.$
This result can also be obtained by expanding $\mathcal{K}_{ab}$
in a series approximation. 

In order to estimate $\mathcal{K}_{ab},$ we set $\Delta_{a}=\Delta_{b}\equiv\Delta$
for simplicity. Since we choose the symmetric operation points $\epsilon_{a}=\epsilon_{b}=0$
for both RX qubits, we also take $t_{a}=t_{b}=t_{c}$ (where $t_{c}\equiv t_{\alpha l}=t_{\alpha r}$
for each qubit $\alpha$ as described in Appendix \ref{sec:RXqubitbas}
and Sec. \ref{subsec:EffHamiltonian}) so that $q_{z}^{a}=q_{z}^{b}.$
The RX qubit parameter values we use (for $\hbar=1$) are $\Delta=2\pi\times30\ {\rm GHz}$
and $t_{c}=2\pi\times14\ {\rm GHz},$ which give $q_{z}^{a}=q_{z}^{b}\approx0.022.$
We also set the mediator dot electric field drive amplitude to be
$\mathcal{E}_{M}=2\ {\rm V/m}.$ The resulting qubit-qubit coupling
strength $\mathcal{K}_{ab}=\left(\Omega_{s}-\Omega_{d}\right)/4$
is plotted in Fig. \ref{fig:couplingstrength} as a function of the
mediator dot size $\lambda$ for multiple values of the distance $a$
between the centers of the mediator dot and the nearest-neighbor dot
within each RX qubit (dot $a3$ or $b1$ in Fig. \ref{fig:RX-MEQD}).
We note that larger values of $\mathcal{K}_{ab}$ can be obtained
by increasing $\lambda$ or decreasing $a$ while satisfying the condition
$\lambda/2a\ll1$ to maintain the validity of Eqs. (\ref{eq:1byr})-(\ref{eq:deltaK}).
In practice, the qubit-qubit coupling strength is bounded from above
by experimentally achievable values of $K_{1}$ and $K_{2}.$ Recent
experiments suggest that interdot capacitive coupling strengths $\sim10-100\ {\rm GHz}$
are feasible \cite{Neyens2019,WangW2024}. 

To quantitatively analyze the entangling gate performance as discussed
in Sec. \ref{sec:gateperf}, we set $\lambda=200\ {\rm nm}$ and $a=500\ {\rm nm}$
such that $\lambda/2a=0.2.$ These values give $K_{1}\approx2\pi\times64\ {\rm GHz},$
$K_{2}\approx2\pi\times69\ {\rm GHz},$ and the mediator dot drive
Rabi frequency $\Omega_{M}\approx2\pi\times97\ {\rm MHz}.$ From these
values, we find $\Delta K\approx2\pi\times4.8\ {\rm GHz}$ and $\mathcal{K}_{ab}\approx2\pi\times34\ {\rm MHz}.$
As discussed in Sec. \ref{sec:gateperf}, this coupling strength gives
rise to rapid two-qubit entangling gates. 

\section{\label{sec:gateperf}Performance of driven dot-mediated entangling
gates}

To evaluate the performance of the two-qubit entangling gate $U_{xx}$
{[}Eq. (\ref{eq:Uxx}){]} generated by the effective driven dot-mediated
interaction between dressed RX qubits given by $H_{-}$ {[}Eq. (\ref{eq:Hminus}){]},
we use a master equation analysis to calculate the fidelity of the
gate in the presence of qubit and mediator dot decoherence. Here,
we consider dephasing for the qubits as well as the mediator dot,
which is expected to be the dominant type of noise for silicon-based
implementations of both systems \cite{Gamble2012,Srinivasa2016,Weinstein2023}. 

We start with the master equation \cite{Srinivasa2016,Srinivasa2024}
\begin{align}
\dot{\rho} & =-i\left[H_{l}^{{\rm eff}},\rho\right]+\sum_{\alpha=a,b}\frac{\gamma_{\alpha}}{2}\left(\sigma_{\alpha}^{z}\rho\sigma_{\alpha}^{z}-\rho\right)\nonumber \\
 & +\frac{\gamma_{M}}{2}\left(\tau_{z}\rho\tau_{z}-\rho\right),\label{eq:mastereq}
\end{align}
where $H_{l}^{{\rm eff}}$ is given in Eq. (\ref{eq:Hleffshiftfreq}),
$\gamma_{\alpha}$ represents the dephasing rate for qubit $\alpha$
in the original basis, and $\gamma_{M}$ represents the mediator dot
dephasing rate. In order to describe the dynamics generated by the
effective dot-mediated two-qubit interaction in the presence of dephasing,
we apply the same series of transformations described in Sec. \ref{subsec:EffHamiltonian}
to the master equation. Moving to a frame rotating at the mediator
dot drive frequency $\omega_{M}^{d}$ via Eq. (\ref{eq:Urf}), considering
the resonantly driven dot case $\omega_{M}^{d}=\omega_{s}^{\prime},$
rotating the driven mediator dot to a dressed singlet representation
via $U_{s}$ {[}Eq. (\ref{eq:Us}){]}, applying a rotating wave approximation
for $\Omega_{M}\ll\left|\delta_{\alpha}\right|,\left|\delta_{\alpha}\pm\Omega_{M}\right|$
in an interaction picture, and dropping rapidly oscillating terms
leads to
\begin{align}
\dot{\rho}_{r} & =-i\left[H_{l,{\rm RWA}}^{{\rm eff}},\rho_{r}\right]+\sum_{\alpha}\frac{\gamma_{\alpha}}{2}\left(\sigma_{\alpha}^{z}\rho_{r}\sigma_{\alpha}^{z}-\rho_{r}\right)\nonumber \\
 & +\frac{\gamma_{M}}{2}\left(\tilde{\tau}_{x}\rho_{r}\tilde{\tau}_{x}-\rho_{r}\right),\label{eq:meqrf}
\end{align}
where $H_{l,{\rm RWA}}^{{\rm eff}}$ is given in Eq. (\ref{eq:HleffRWA})
and $\rho_{r}\equiv U_{s}^{\dagger}U_{{\rm rf}}^{\dagger}\rho U_{{\rm rf}}U_{s}$
after applying the same rotating wave approximations leading to $H_{l,{\rm RWA}}^{{\rm eff}}.$ 

We next diagonalize the mediator dot subspace via $U_{d}$ {[}Eq.
(\ref{eq:Ud}){]}, which yields 
\begin{align}
\dot{\rho}_{d} & =-i\left[H_{l,d}^{{\rm eff}},\rho_{d}\right]+\sum_{\alpha}\frac{\gamma_{\alpha}}{2}\left(\sigma_{\alpha}^{z}\rho_{d}\sigma_{\alpha}^{z}-\rho_{d}\right)\nonumber \\
 & +\frac{\gamma_{M}}{2}\left(T_{x}\rho_{d}T_{x}-\rho_{d}\right)\label{eq:meqd}
\end{align}
with $H_{l,d}^{{\rm eff}}$ given in Eq. (\ref{eq:Hleffd}) and $\rho_{d}\equiv U_{d}^{\dagger}\rho_{{\rm RWA}}U_{d}.$
In writing Eq. (\ref{eq:meqd}), we have used the fact that $\left[U_{d},\sigma_{\alpha}^{z}\right]=0$
and have also defined the operator
\begin{align}
T_{x} & \equiv U_{d}^{\dagger}\tilde{\tau}_{x}U_{d}\nonumber \\
 & =\frac{1}{2}\left[\left(\cos\theta_{s}+\cos\theta_{d}\right){\bf 1}+\left(\cos\theta_{s}-\cos\theta_{d}\right)\sigma_{a}^{z}\sigma_{b}^{z}\right]\tilde{\tau}_{x}\nonumber \\
 & +\frac{1}{2}\left[\left(\sin\theta_{s}+\sin\theta_{d}\right)\sigma_{a}^{z}+\left(\sin\theta_{s}-\sin\theta_{d}\right)\sigma_{b}^{z}\right]\tilde{\tau}_{z}.\label{eq:Tx}
\end{align}
We note that $T_{x}$ acts in the full space consisting of both qubits
and the mediator dot, such that applying $U_{d}$ serves to translate
the effect of the mediator dot decay into the two-qubit space due
to the capacitive coupling. Transforming to the dressed basis for
RX qubits chosen in this work via $U_{q}$ {[}Eq. (\ref{eq:Uq}){]}
then leads to
\begin{align}
\dot{\rho}_{q} & =-i\left[H_{q},\rho_{q}\right]+\sum_{\alpha}\frac{\gamma_{\alpha}}{2}\left(\tilde{\sigma}_{a}^{x}\rho_{q}\tilde{\sigma}_{a}^{x}-\rho_{q}\right)\nonumber \\
 & +\frac{\gamma_{M}}{2}\left(\tilde{T}_{x}\rho_{q}\tilde{T}_{x}-\rho_{q}\right)\label{eq:meqq}
\end{align}
where $H_{q}$ is given in Eq. (\ref{eq:Hq}), $\rho_{q}\equiv U_{q}^{\dagger}\rho_{d}U_{q},$
and 
\begin{align}
\tilde{T}_{x} & \equiv U_{q}^{\dagger}T_{x}U_{q}\nonumber \\
 & =\frac{1}{2}\left[\left(\cos\theta_{s}+\cos\theta_{d}\right){\bf 1}+\left(\cos\theta_{s}-\cos\theta_{d}\right)\tilde{\sigma}_{a}^{x}\tilde{\sigma}_{b}^{x}\right]\tilde{\tau}_{x}\nonumber \\
 & -\frac{1}{2}\left[\left(\sin\theta_{s}+\sin\theta_{d}\right)\tilde{\sigma}_{a}^{x}+\left(\sin\theta_{s}-\sin\theta_{d}\right)\tilde{\sigma}_{b}^{x}\right]\tilde{\tau}_{z}.\label{eq:Txtilde}
\end{align}
Finally, to focus on the slow dynamics due to just the interaction
term in $H_{q}$, we transform to an interaction picture by writing
$H_{q}=H_{0,q}+V_{q},$ where 
\begin{align}
H_{0,q} & \equiv-\sum_{\alpha=a,b}\frac{\delta_{\alpha}}{2}\tilde{\sigma}_{\alpha}^{x}+\frac{\Omega_{M}^{\prime}}{2}\tau_{z}^{M},\nonumber \\
V_{q} & \equiv\mathcal{K}_{ab}\tilde{\sigma}_{a}^{x}\tilde{\sigma}_{b}^{x}\tau_{z}^{M},\label{eq:H0qVq}
\end{align}
and using $U_{{\rm int}}=e^{-iH_{0,q}t}.$ In this interaction picture,
we find $V^{\prime}\equiv U_{{\rm int}}^{\dagger}H_{q}U_{{\rm int}}-iU_{{\rm int}}^{\dagger}\dot{U}_{{\rm int}}=V_{q}$
and
\begin{align}
\dot{\rho}^{\prime} & =-i\left[V^{\prime},\rho^{\prime}\right]+\sum_{\alpha}\frac{\gamma_{\alpha}}{2}\left(\tilde{\sigma}_{a}^{x}\rho^{\prime}\tilde{\sigma}_{a}^{x}-\rho^{\prime}\right)\nonumber \\
 & +\frac{\gamma_{M}}{2}\left(T_{x}^{\prime}\rho^{\prime}T_{x}^{\prime}-\rho^{\prime}\right),\label{eq:meqpr}
\end{align}
where $\rho^{\prime}\equiv U_{{\rm int}}^{\dagger}\rho_{q}U_{{\rm int}}$
and $T_{x}^{\prime}\equiv U_{{\rm int}}^{\dagger}\tilde{T}_{x}U_{{\rm int}}.$
Note that taking $\tau_{z}^{M}\rightarrow-1$ in $V^{\prime}$ yields
the two-qubit interaction term $V_{-}\equiv-\mathcal{K}_{ab}\tilde{\sigma}_{a}^{x}\tilde{\sigma}_{b}^{x}$
in $H_{-}.$ 

Using the solution to Eq. (\ref{eq:meqpr}), we evaluate the performance
of the two-qubit gate in Eq. (\ref{eq:Uxx}) in the presence of decay
via the fidelity \cite{Vandersypen2005,Srinivasa2024}
\begin{equation}
F\left(t_{g}\right)\equiv{\rm Tr}\left[\rho^{\prime{\rm \left(0\right)}}\left(t_{g}\right)\rho^{\prime}\left(t_{g}\right)\right],\label{eq:F}
\end{equation}
where $\rho^{\prime}\left(t_{g}\right)$ denotes the final state at
time $t_{g}$ for the evolution obtained via numerical integration
of Eq. (\ref{eq:meqpr}) and $\rho^{\prime{\rm \left(0\right)}}\left(t_{g}\right)$
denotes the final state for the ideal evolution given by setting $\gamma_{\alpha}=\gamma_{M}=0.$
We note that, due to the block-diagonal structure of $V^{\prime},$
the ideal evolution in the interaction picture is given exactly by
$U_{-}^{\prime}\left(t\right)=e^{-iV_{-}t}$ within the $\tau_{z}^{M}\rightarrow-1$
subspace. For $t=t_{g}=\pi/4\mathcal{K}_{ab}$ (i.e., setting $m=0$)
and $\delta_{a}=\delta_{b}=8r\mathcal{K}_{ab}$ with integer $r,$
$U_{-}^{\prime}\left(t_{g}\right)=U_{-}\left(t_{g}\right)=U_{xx}$
as given in Eq. (\ref{eq:Uxx}). 

\begin{figure}
\begin{centering}
\includegraphics[viewport=0bp 40bp 660bp 540bp,width=0.45\textwidth]{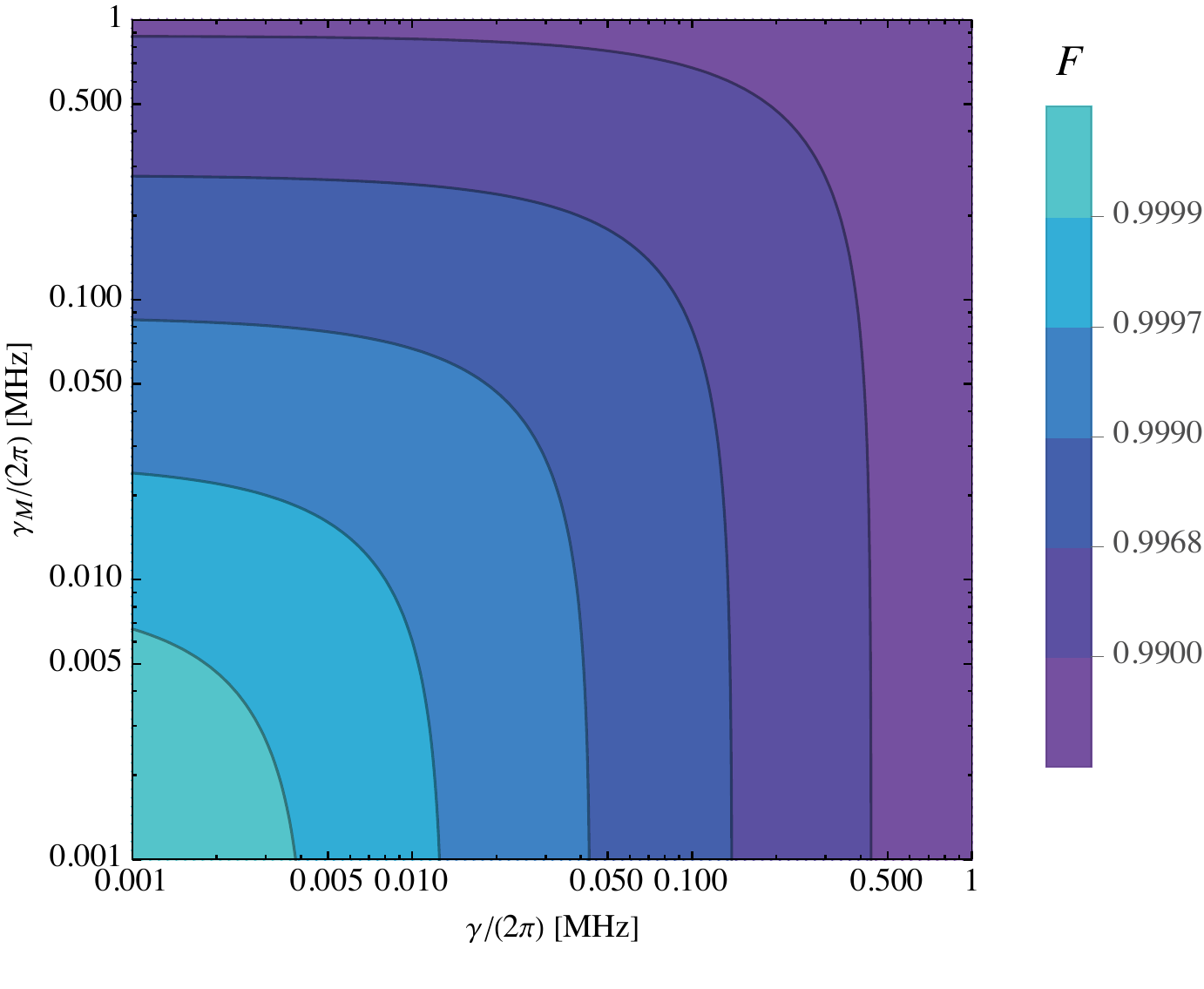}
\par\end{centering}
\caption{\label{fig:Fvsdephrates}Fidelity $F$ for the driven dot-mediated
gate generated by the interaction $V^{\prime}$ according to Eq. (\ref{eq:meqpr})
as a function of the qubit decay rate $\gamma$ and the mediator dot
decay rate $\gamma_{M},$ calculated using Eq. (\ref{eq:F}) with
the initial state $\ket{\psi_{i}}=\ket{eg,M_{-}}.$ The ideal evolution
is given by $U_{xx}$ {[}Eq. (\ref{eq:Uxx}){]}, which is equivalent
to the action of $U_{i{\rm SW}}^{1/2}$ within the $\left\{ \ket{eg,M_{-}},\ket{ge,M_{-}}\right\} $
subspace.}
\end{figure}

As a concrete illustration, we focus on the gate $U_{i{\rm SW}}^{1/2}$
and accordingly take the initial state to be $\rho^{\prime}\left(0\right)=\ket{\psi_{i}}\bra{\psi_{i}},$
where $\ket{\psi_{i}}=\ket{eg,M_{-}}$ denotes the state of the full
qubit-mediator dot system associated with the dressed two-qubit state
$\ket{eg}$ in the $\tau_{z}^{M}\rightarrow-1$ mediator dot subspace.
In this case, the ideal final state is given by 
\begin{align}
\rho^{\prime{\rm \left(0\right)}}\left(t_{g}\right) & =U_{-}\left(t_{g}\right)\rho^{\prime}\left(0\right)U_{-}^{\dagger}\left(t_{g}\right)\nonumber \\
 & =U_{xx}\rho^{\prime}\left(0\right)U_{xx}^{\dagger}\nonumber \\
 & =\ket{\psi_{f}}\bra{\psi_{f}}\label{eq:rhotgid}
\end{align}
with $\ket{\psi_{f}}=\left(\ket{eg,M_{-}}+i\ket{ge,M_{-}}\right)/\sqrt{2}.$
The analysis for the gate $U_{i{\rm DE}}^{1/2}$ with the initial
state $\ket{\psi_{i}}=\ket{ee,M_{-}}$ is related to that for the
gate $U_{i{\rm SW}}^{1/2}$ with $\ket{\psi_{i}}=\ket{eg,M_{-}}$
via a unitary rotation of the second qubit and thus yields analogous
results. 
\begin{center}
\begin{figure*}
\begin{centering}
\includegraphics[viewport=80bp 20bp 915bp 220bp,width=0.6\paperwidth]{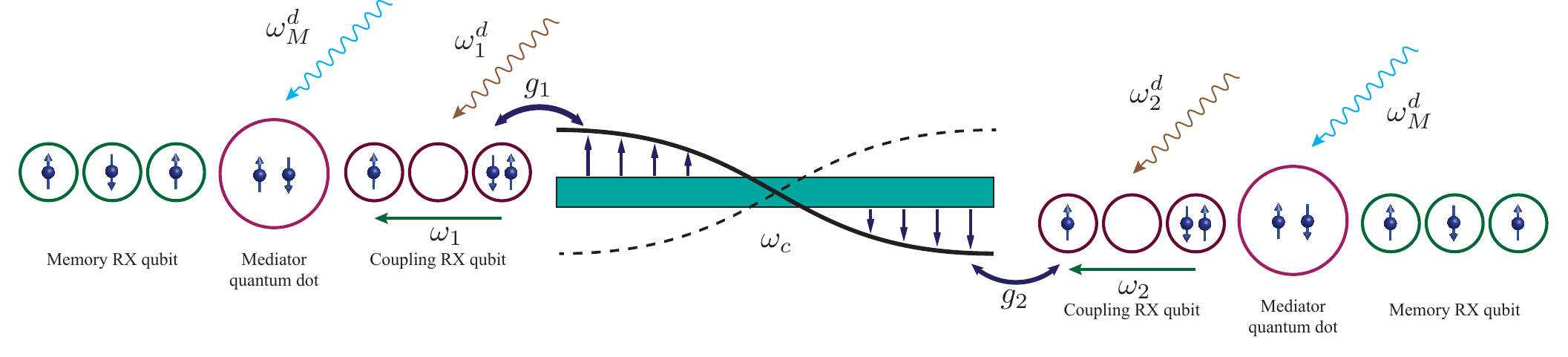}
\par\end{centering}
\caption{\label{fig:RX-MEQD-modularsys} Illustration of the building blocks
of a modular system envisioned for spin-based quantum information
processing based on two drive-switchable entanglement modes, consisting
of (1) intramodular entangling interactions activated by electrically
driving the mediator dots (blue wave arrows) and (2) intermodular
entangling interactions via sidebands activated by driving the qubits
coupled to the cavity (brown wave arrows).}
\end{figure*}
\par\end{center}

In order to calculate the fidelity in Eq. (\ref{eq:F}), we solve
the master equation in Eq. (\ref{eq:meqpr}) for $\rho^{\prime}\left(t_{g}\right)$
via numerical integration with the parameters specified in Sec. \ref{subsec:parameters_Kab}
and Appendix \ref{sec:RWAvalidity}, which give a qubit-qubit coupling
strength $\mathcal{K}_{ab}\approx2\pi\times34\ {\rm MHz}.$ These
parameters yield a two-qubit gate time $t_{g}\approx3.7\ {\rm ns}.$
Within a qubit dephasing time $T_{2}^{\ast}=1/\gamma=3.5\ \mu{\rm s}$
\cite{Weinstein2023}, this gate time corresponds to $T_{2}^{\ast}/t_{g}\approx950$
two-qubit gate operations. We plot the fidelity given by Eq. (\ref{eq:F})
for the gate $U_{i{\rm SW}}^{1/2}$ as a function of the qubit and
mediator dot dephasing rates $\gamma\equiv\gamma_{a}=\gamma_{b}$
and $\gamma_{M},$ where we take the dephasing rates of the two RX
qubits to be equal for simplicity (Fig. \ref{fig:Fvsdephrates}).
We find fidelities $F>0.99$ for $\gamma\lesssim2\pi\times0.43\ {\rm MHz}$
or for $\gamma_{M}\lesssim2\pi\times0.87\ {\rm MHz}.$ We also note
that the fidelity is less sensitive to the mediator dot dephasing
compared to the qubit dephasing, despite the mainly charge dipole
character of the singlet-singlet transition mediating the qubit-qubit
interaction. This reduced sensitivity is consistent with the origin
of $\Delta K$ from the quadrupole terms in the Coulomb interaction
noted in Sec. \ref{subsec:parameters_Kab}. Additionally, the condition
$\gamma_{M}\ll\Omega_{M}$ for suppression of dephasing in the dressed
mediator dot basis is satisfied for the Rabi frequency $\Omega_{M}\approx2\pi\times97\ {\rm MHz}$
calculated in Sec. \ref{subsec:parameters_Kab} and the range of $\gamma_{M}$
values in Fig. \ref{fig:Fvsdephrates}. 

As a measure of the strength of the coherent coupling relative to
the dephasing rates, we can calculate a quantity analogous to the
cooperativity for qubit-cavity photon interaction \cite{Childress2004,Blais2004}
that we define as $\mathcal{C}\equiv\mathcal{K}_{ab}^{2}/\gamma_{M}\gamma.$
For $\gamma=2\pi\times0.25\ {\rm MHz}$ and $\gamma_{M}=2\pi\times0.37\ {\rm MHz},$
which together yield $F>0.99,$ we find $\mathcal{C}\approx1.3\times10^{4}.$
Finally, a numerical analysis using a 32-dimensional Hamiltonian matrix
for $H_{l}$ {[}Eq. \ref{eq:Hl}{]} that includes the higher-energy
states $\ket{+}_{g}$ and $\ket{+}_{e}$ for each qubit shows that
the leakage out of the two-qubit subspace, given by \cite{Wardrop2016}
$\mathcal{L}\equiv{\rm Tr}\left[Q\rho\left(t\right)Q\right]$ with
$\rho\left(t\right)$ denoting the solution to the master equation
in Eq. (\ref{eq:mastereq}) with the qubit operators replaced by their
four-dimensional counterparts and $Q\equiv1-P$ with $P$ the projector
onto the eight-dimensional subspace in which $H_{l}^{{\rm eff}}$
{[}Eq. \ref{eq:Hleff}{]} is defined, remains bounded such that $\mathcal{L}<0.13$
for all times up to the maximum time $t=9t_{g}$ used in the calculation.
As seen from the analysis in Sec. \ref{subsec:parameters_Kab}, the
coupling strength $\mathcal{K}_{ab}$ that we use in these calculations
does not represent a fundamental upper limit, and larger experimentally
achievable values of $\mathcal{K}_{ab}$ can be used to achieve higher
fidelities. 

\section{\label{sec:spinqubitmodularity}Outlook: Spin qubit modularity via
drive-activated entanglement}

As discussed in Sec. \ref{sec:Theoretical-framework} and Appendix
\ref{sec:MEQDdipole}, the capacitive spin qubit entangling interaction
that we present in this work is activated via the drive on the two-electron
mediator dot and does not exist in the absence of the drive. The switchable
nature of the interaction via the driven dot in principle enables
the integration of the intramodular entangling approach in this work
with the microwave cavity photon-mediated intermodular entangling
approach for spin qubits described in Ref. \cite{Srinivasa2024},
where driving of the qubits enables tunable and spectrally flexible
long-range entangling gates via the generated qubit sidebands. The
fact that the sideband-based intermodular interaction can also be
switched on and off via the drive on the qubits suggests a method
for achieving modularity with spin qubits via two distinct modes of
entanglement {[}Fig. \ref{fig:RX-MEQD-modularsys}{]} that we now
briefly describe. 

In the intramodular entangling mode, the driving fields on the mediator
dots are switched on to generate local entanglement between qubits
within each module that serve as memory qubits, while keeping the
drives on the qubits interacting with the cavity (labeled as coupling
qubits in Fig. \ref{fig:RX-MEQD-modularsys}) switched off to suppress
long-range entanglement between modules. On the other hand, the intermodular
entangling mode involves turning on the drives on the coupling qubits
to generate entanglement via sideband resonances as specified in Ref.
\cite{Srinivasa2024} while switching off the drives on the mediator
dots to suppress local entanglement between the coupling and memory
qubits. These distinct intramodular and intermodular entangling modes
are enabled by the off-resonant coupling of the driven qubits to the
cavity via sidebands with $\omega_{1}\neq\omega_{2}\neq\omega_{c},$
where $\omega_{1}$ and $\omega_{2}$ are the frequencies of the qubits
(denoted by $\omega_{\alpha}$ for $\alpha=a,b$ in this work) coupled
to the cavity and $\omega_{c}$ is the cavity frequency \cite{Srinivasa2024},
together with the drive-activated nature of both entangling approaches.
Switching between the intermodular and intramodular coupling regimes
can be carried out via switching between the ac driving fields on
the qubits and those on the mediator dots. Cross-talk between intramodular
and intermodular entanglement can also in principle be suppressed
via the separation of energy scales that exists as a consequence of
the fact that $\omega_{s}\gg\omega_{\alpha}$ for typical systems
(see Appendix \ref{sec:RWAvalidity}), as well as by virtue of the
spatial separation provided by both the cavity \cite{Childress2004}
and the mediator dots \cite{Beil2014thesis,Croot2018,Malinowski2019,Fedele2021}. 

A variety of potential future directions may be envisioned for the
work presented here. Follow-up work could involve investigating alternative
geometries for driven dot-mediated coupling between two RX qubits
\cite{Taylor2013,Doherty2013,Wardrop2016,Pal2014}, as well as adapting
the general theory to other types of spin qubits that can be capacitively
coupled as noted previously to determine parameter regimes required
for entangling gates, including those for suppressing potential singlet-triplet
coupling induced by spin-charge mixing mechanisms. Another direction
of interest involves investigating an extension of the spin qubit
entangling approach to mediator dots with more than two electrons,
which may enable improved entangling fidelities via screening of charge
impurities \cite{Vorojtsov2004,Barnes2011,Higginbotham2014,Malinowski2018,Malinowski2019,Fedele2021,PaqueletWuetz2023}. 

The finite transition dipole moment of the driven two-electron mediator
dot, which originates from the distinct charge distributions associated
with the mediator dot orbital levels $c1$ and $c2$ (see Appendix
\ref{sec:MEQDdipole}) and which allows for capacitive interaction
between the RX qubits and the mediator dot through $\Delta K\neq0$
{[}see Eq. (\ref{eq:HleffRWA}) and Sec. \ref{subsec:parameters_Kab}{]},
also suggests the possibilities of dispersive coupling of the driven
mediator dot to a microwave cavity and cavity-based dispersive readout
\cite{Blais2004} of the $2\rightarrow2^{\ast}$ mediator dot charge
transition. Quantitatively, since $\omega_{c}<2\pi\times10\ {\rm GHz}$
for typical circuit quantum electrodynamics experiments with spin
qubits \cite{Landig2018,Landig2019,Borjans2020,Harvey-Collard2022,Yu2023,Dijkema2025,Jiang2025},
the physically relevant case corresponds to $\omega_{s}>\omega_{c}$
(see Appendix \ref{sec:RWAvalidity}) and thus the dispersive regime
of mediator dot-resonator interaction for a dot-photon coupling strength
that is small compared to the detuning $\omega_{s}-\omega_{c}.$ Resonator-mediated
interaction between pairs of driven mediator dots that serve as more
noise-resilient interface qubits for coupling to photons or the coupling
of multiple spin qubits via a common driven mediator dot \cite{Fedele2021}
may also prove useful for extension of the modular system to higher
dimensions and for distributing entanglement throughout a modular
spin qubit network. 

\section{Conclusions}

In this work, we have developed an approach for achieving a tunable
interaction and rapid entangling gates between a pair of spin qubits
via capacitive coupling mediated by an ac-driven two-electron mediator
quantum dot. The entangling interaction is activated via the driving
field applied to the mediator dot and therefore can be switched on
and off via this drive. By tailoring the spectral properties of the
mediator dot, the drive also serves to simplify both the low-energy
description of the mediator dot and the form of the qubit-qubit interaction.
The resulting coupling generates rapid, single-pulse universal entangling
gates between RX qubits with expected few-nanosecond gate rates that
are comparable to those of state-of-the-art exchange-based two-qubit
gates \cite{He2019,Hendrickx2020,Noiri2022,Xue2022,Mills2022,Weinstein2023,WangCA2024,Madzik2025}
and significantly faster than demonstrated capacitive coupling-based
two-qubit gates \cite{Shulman2012,Nichol2017}. As the underlying
coupling mechanism is capacitive and the driven mediator dot can be
described as an effective two-level singlet system, this gate conserves
the spin of individual qubits and thus does not require the extensive
pulse sequence typically needed to mitigate leakage in exchange interaction-based
two-qubit gates between conventional exchange-only spin qubits. The
general formalism can also be adapted to a wide variety of qubit encodings
with spin-dependent charge states that enable capacitive interaction.

In the dressed qubit basis, we find that the driven dot-mediated local
entangling gates are of the same types as those derived for the long-range
sideband-based entangling gates mediated by microwave cavity photons
between driven qubits in Ref. \cite{Srinivasa2024}. The drive-activated
character of both local entanglement within spin qubit modules and
long-distance entangling interactions between modules enables the
identification of two distinct coupling regimes: (1) an intramodular
capacitive coupling regime, in which mediator dots are driven to generate
entanglement with the RX qubit drives off; and (2) an intermodular
spin-photon coupling regime, in which RX qubits coupled to cavity
photons are driven to generate entanglement via sidebands with the
mediator dot drives off. The ability to switch between these two entangling
regimes and the flexibility enabled by drive-activated control in
principle allows for integration of intramodular and intermodular
entanglement, establishing the combined system as a promising building
block for full modularity in spin-based quantum information processing. 
\begin{acknowledgments}
We thank J. M. Taylor, C. M. Marcus, F. Kuemmeth, and S. Coppersmith
for helpful discussions. This work was supported by Army Research
Office Grant W911NF-23-1-0104. 
\end{acknowledgments}

\appendix

\section{\label{sec:MEQDdipole}Driven mediator dot Hamiltonian}

Here, we describe the effective Hamiltonian for the ac-driven two-level,
two-electron mediator quantum dot and derive the form of the electric
dipole interaction that gives rise to Eq. (\ref{eq:HMd}). The electric
dipole operator for a quantum dot with multielectron occupation of
levels $c1$ and $c2$ is given by 

\begin{equation}
\mathbf{d}=\sum_{i,j=1,2}\sum_{\sigma=\uparrow,\downarrow}\bra{ci}\mathbf{d}\ket{cj}c_{ci\sigma}^{\dagger}c_{cj\sigma},\label{eq:dipccdag}
\end{equation}
where $\mathbf{d}=-e\mathbf{r}=-e\left(x\hat{x}+y\hat{y}\right)$
is the dipole operator of a single electron confined to a quantum
dot defined in the $x$-$y$ plane. Here, we take $c1$ and $c2$
to represent orbital levels of the mediator dot. For silicon quantum
dots, in which low-lying valley states may also exist in the conduction
band \cite{Zwanenburg2013}, we assume a large valley splitting energy
$E_{{\rm V}}\gtrsim100\ \mu{\rm eV}$ \cite{Zwanenburg2013,Yang2013,Hollmann2020,McJunkin2022,DegliEsposti2024,Stehouwer2025}
and consider only the lowest-energy valley state for the electrons.
While a transition dipole moment is also possible between valley states,
this dipole moment is typically much smaller than the orbital electric
dipole moment \cite{Gamble2012,Yang2013,Gamble2013,Salamone2026}.
We also consider a circular dot, for which which Wigner molecularization
effects that may reduce the distinction between the singlet and triplet
charge density distributions due to electron-electron interactions
are expected to be minimized \cite{AbadilloUriel2021Wigner,Ercan2021}.
Accordingly, we evaluate the one-electron matrix elements in Eq. (\ref{eq:dipccdag})
using the Fock-Darwin wave functions for the ground and first-excited
orbital levels of the mediator dot in the presence of a perpendicular
magnetic field $\mathbf{B}=B\hat{z}$ \cite{Kouwenhoven2001,Yang2011Hubbard,Yang2011geometry,Gamble2012},
\begin{align}
\psi_{c1}\left(x,y\right) & =\frac{e^{-\left(x^{2}+y^{2}\right)/4\lambda^{2}}}{\sqrt{2\pi}\lambda},\nonumber \\
\psi_{c2}\left(x,y\right) & =\frac{\left(x+iy\right)e^{-\left(x^{2}+y^{2}\right)/4\lambda^{2}}}{2\sqrt{\pi}\lambda^{2}},\label{eq:psic1c2}
\end{align}
which define the mediator dot size $\lambda.$ The integrals involved
in calculating the matrix elements can be evaluated by changing to
polar coordinates $\left(r,\theta\right),$ where $r^{2}=x^{2}+y^{2}$
and $\tan\theta=y/x.$ We find
\begin{align}
\bra{c2}x\ket{c1} & =\frac{\lambda}{\sqrt{2}},\label{eq:dipxmatel}\\
\bra{c2}y\ket{c1} & =-\frac{i\lambda}{\sqrt{2}},\label{eq:dipymatel}
\end{align}
while parity considerations show that the diagonal matrix elements
of the dipole operator vanish. Substituting the one-electron dipole
operator $\mathbf{d}$ into Eq. (\ref{eq:dipccdag}) and using the
evaluated matrix elements yields the multielectron quantum dot dipole
operator expression. 

In this work, we assume for simplicity that the driving field applied
to the mediator dot is directed along the $x$ axis such that $\mathcal{\boldsymbol{E}}=\mathcal{E}_{x}\hat{x},$
where $\mathcal{E}_{x}=\mathcal{E}_{0}+\mathcal{E}_{M}\cos\left(\omega_{M}^{d}t+\phi_{M}\right)$
with $\mathcal{E}_{0}$ denoting the dc component and $\mathcal{E}_{M},$
$\omega_{M}^{d},$ and $\phi_{M}$ denoting the amplitude, frequency,
and phase, respectively, of the ac component of the driving field.
The dipole interaction then becomes $H_{{\rm dip}}=-\mathbf{d}\cdot\boldsymbol{\mathcal{E}}=-d_{x}\mathcal{E}_{x}=ex\mathcal{E}_{x},$
where we can use Eqs. (\ref{eq:dipccdag}) and (\ref{eq:dipxmatel})
to write the dipole operator as
\begin{equation}
d_{x}=-\frac{e\lambda}{\sqrt{2}}\sum_{\sigma=\uparrow,\downarrow}\left(c_{c1\sigma}^{\dagger}c_{c2\sigma}+c_{c2\sigma}^{\dagger}c_{c1\sigma}\right).\label{eq:dipx}
\end{equation}
From Eq. (\ref{eq:dipx}), we see that the electric dipole describes
the spin-conserving transfer of electrons between the two levels of
the mediator dot and scales with the dot size $\lambda.$ 

The low-energy spectrum of a two-level, two-electron dot has been
investigated in detail in Ref. \cite{Srinivasa2015}. Here, we consider
the same set of five low-lying states to describe the mediator dot,
consisting of the doubly occupied lower-energy singlet state $\ket{S_{11}},$
the singly occupied higher-energy singlet state $\ket{S_{12}},$ and
the singly occupied triplet states $\ket{T_{12}^{\left(-\right)}},$
$\ket{T_{12}^{\left(0\right)}},$ and $\ket{T_{12}^{\left(+\right)}}$
as illustrated in Fig. \ref{fig:mediatordotspectrum}(a) (see also
Fig. 1(c) of Ref. \cite{Srinivasa2015}). Within the low-energy subspace
spanned by these states, we write the mediator dot Hamiltonian $H_{c}$
{[}Eq. (\ref{eq:Hc}){]} as
\begin{equation}
H_{c}^{\prime}=\omega_{s}\ket{S_{12}}\bra{S_{12}}+\Delta_{T}P_{T}.\label{eq:Hcpr}
\end{equation}
In Eq. (\ref{eq:Hcpr}), $P_{T}\equiv\ket{T_{12}^{\left(-\right)}}\bra{T_{12}^{\left(-\right)}}+\ket{T_{12}^{\left(0\right)}}\bra{T_{12}^{\left(0\right)}}+\ket{T_{12}^{\left(+\right)}}\bra{T_{12}^{\left(+\right)}}$
is the projector onto the triplet subspace, and we have chosen the
energy origin to be the energy of the two-electron ground state $\ket{S_{11}},$
given by $E\left(S_{11}\right)\equiv-2\epsilon_{c1}+U_{c}.$ We have
also defined the energy splitting between the two singlet states $\ket{S_{11}}$
and $\ket{S_{12}}$ as $\omega_{s}\equiv-\left(\epsilon_{c2}-\epsilon_{c1}\right)-U_{c}+K_{c}+J_{c}$
and that between $\ket{S_{11}}$ and the three triplet states in the
absence of a magnetic field as $\Delta_{T}\equiv\omega_{s}-2J_{c}$
(we note that this definition of the gap between $\ket{S_{11}}$ and
the triplet states is different from the corresponding definition
of $\Delta_{M}$ used in Ref. \cite{Srinivasa2015}, which includes
additional terms involving the Coulomb interaction of the center dot
electrons with electrons external to the center dot). 

In the two-electron, five-state representation used to write Eq. (\ref{eq:Hcpr}),
we find that the electric dipole operator in Eq. (\ref{eq:dipx})
takes the form
\begin{equation}
d_{x}=-e\lambda\left(\ket{S_{12}}\bra{S_{11}}+\ket{S_{11}}\bra{S_{12}}\right).\label{eq:dipxSSbas}
\end{equation}
From this form of $d_{x},$ we see that the electric dipole operator
for the mediator dot is expressed in terms of the singlet states only
and does not involve the triplet states. The dipole interaction $H_{{\rm dip}}=-d_{x}\mathcal{E}_{x}$
therefore couples only the singlet states, while all triplet states
are decoupled from the singlet subspace and are not excited via the
electric field as we describe in more detail below. 

In general, the Hamiltonian of the mediator dot in Eq. (\ref{eq:Hc})
will contain an additional term of the form 
\begin{equation}
H_{u}=u\sum_{\sigma\neq\sigma^{\prime}}\left(n_{c1\sigma}+n_{c2\sigma}\right)\left(c_{c1\sigma^{\prime}}^{\dagger}c_{c2\sigma^{\prime}}+{\rm H.c.}\right)\label{eq:Hu}
\end{equation}
that describes on-site occupation-modulated hopping of electrons between
the orbitals of the mediator dot \cite{Hubbard1963,Yang2011Hubbard,Yang2011geometry}.
In the low-energy basis of the dot, $H_{u}$ leads to a term $\sqrt{2}u\left(\ket{S_{12}}\bra{S_{11}}+\ket{S_{11}}\bra{S_{12}}\right)$
proportional to the dipole operator {[}compare Eq. (\ref{eq:dipxSSbas}){]}
and therefore acts as a dc shift of the dipole interaction for an
electric field applied to the dot. Here, we assume for simplicity
that the dc component $\mathcal{E}_{0}$ of the electric field $\mathcal{E}_{x}$
is chosen to offset the effect of $H_{u}$ such that $\mathcal{E}_{0}=-\sqrt{2}u/e\lambda$
and consider only the ac component of the electric field. Using Eq.
(\ref{eq:dipxSSbas}), the dipole interaction $H_{{\rm dip}}$ then
becomes
\begin{align}
H_{M}^{d} & \equiv\Omega_{M}\cos\left(\omega_{M}^{d}t+\phi_{M}\right)\left(\ket{S_{12}}\bra{S_{11}}+\ket{S_{11}}\bra{S_{12}}\right),\label{eq:Hdip-HMd}
\end{align}
where we have defined the Rabi frequency $\Omega_{M}\equiv e\lambda\mathcal{E}_{M}.$
We thus obtain Eq. (\ref{eq:HMd}) in the main text. 

The form of the electric dipole interaction in Eq. (\ref{eq:Hdip-HMd})
highlights the decoupled nature of the low-energy singlet and triplet
subspaces of the driven two-electron mediator dot. This decoupling
gives rise to an effective two-level description in terms of field-dressed
singlet states, as shown in Fig. \ref{fig:mediatordotspectrum}(b).
To illustrate this description more explicitly, we use Eqs. (\ref{eq:Hcpr})
and (\ref{eq:Hdip-HMd}) to write the full mediator dot Hamiltonian
as $H_{M}^{\prime}\equiv H_{c}^{\prime}+H_{M}^{d}$ and then transform
$H_{M}^{\prime}$ to a frame rotating at the drive frequency $\omega_{M}^{d}$
via
\begin{equation}
U_{M}=e^{-i\omega_{M}^{d}t\left(\ket{S_{12}}\bra{S_{12}}+P_{T}\right)},\label{eq:UM}
\end{equation}
such that $H_{M}^{{\rm rf}}\equiv U_{M}^{\dagger}H_{M}^{\prime}U_{M}-iU_{M}^{\dagger}\dot{U}_{M}.$
Setting $\phi_{M}=0$ for simplicity, defining the mediator dot-drive
detuning $\Delta_{s}\equiv\omega_{s}-\omega_{M}^{d},$ making a rotating
wave approximation for $\left|\Delta_{s}\right|\ll\omega_{M}^{d},$
and dropping rapidly oscillating terms $\sim e^{\pm2i\omega_{M}^{d}t}$
yields
\begin{align}
H_{M}^{{\rm rf}} & \approx H_{M}^{{\rm RWA}}\equiv\Delta_{s}\ket{S_{12}}\bra{S_{12}}\nonumber \\
 & +\frac{\Omega_{M}}{2}\left(\ket{S_{12}}\bra{S_{11}}+\ket{S_{11}}\bra{S_{12}}\right)\nonumber \\
 & +\left(\Delta_{T}-\omega_{M}^{d}\right)P_{T}.\label{eq:HM-RWA}
\end{align}
Assuming for simplicity that the mediator dot is driven on resonance
such that $\Delta_{s}=0$ (here, we consider the uncoupled mediator
dot and accordingly neglect the capacitive interaction terms that
shift $\omega_{s}$ {[}see Eq. (\ref{eq:Hleffshiftfreq}){]}), we
diagonalize $H_{M}^{{\rm RWA}}$ via the rotation $U_{{\rm rot}}=e^{-i\pi\tau_{y}/4}$,
which has an action identical to $U_{s}$ {[}Eq. (\ref{eq:Us}){]}
within the singlet subspace and represents the extension of $U_{s}$
into the full five-state low-energy space of the mediator dot. We
find
\begin{align}
\tilde{H}_{M} & \equiv U_{{\rm rot}}^{\dagger}H_{M}^{{\rm RWA}}U_{{\rm rot}}\nonumber \\
 & =\frac{\Omega_{M}}{2}\tilde{\tau}_{z}+\left(\Delta_{T}-\omega_{M}^{d}\right)P_{T},\label{eq:HMtilde}
\end{align}
where $\tilde{\tau}_{z}$ is the Pauli $z$ operator in the basis
of the dressed singlet states $\ket{m_{\pm}}$ used to write Eq. (\ref{eq:Hleffs}).
Note that the triplet basis states are not dressed by the field and
remain unchanged under the transformation $U_{{\rm rot}}.$ 

The spectrum of $\tilde{H}_{M}$ is shown in Fig. (\ref{fig:mediatordotspectrum})(b)
for the resonantly driven dot with $\omega_{M}^{d}=\omega_{s},$ such
that $\left|\Delta_{T}-\omega_{M}^{d}\right|=\omega_{M}^{d}-\Delta_{T}.$
We see from this diagram that the two-level approximation of the mediator
dot in the rotating frame and dressed basis is valid provided $\Omega_{M}\ll\left|\Delta_{T}-\omega_{M}^{d}\right|.$
We also assume a sufficiently small external magnetic field strength
such that the Zeeman splitting between the triplet states (see Fig.
1(c) in Ref. \cite{Srinivasa2015}) is small relative to the singlet-triplet
gap $\left|\Delta_{T}-\omega_{M}^{d}\right|.$ In the absence of additional
terms in the Hamiltonian that do not conserve spin, and at sufficiently
low temperatures such that thermal excitation energies are small compared
to$\left|\Delta_{T}-\omega_{M}^{d}\right|$, we can thus describe
the driven mediator dot within the two-dimensional singlet subspace
alone. 

Including only the singlet terms in $H_{M}^{\prime}=H_{c}^{\prime}+H_{M}^{d}$
and redefining the energy origin for the mediator dot to be $E_{0}^{c}\equiv E\left(S_{11}\right)+\omega_{s}/2,$
we can write the effective Hamiltonian in the singlet subspace as
$H_{c,s}=\omega_{s}\tau_{z}/2,$ where $\tau_{z}\equiv\ket{S_{12}}\bra{S_{12}}-\ket{S_{11}}\bra{S_{11}}$
is a Pauli $z$ operator within the singlet subspace. Combining $H_{c,s}$
with Eq. (\ref{eq:HMd}), we then find\ref{fig:RX-MEQD}
\begin{equation}
H_{M}^{{\rm eff}}=\frac{\omega_{s}}{2}\tau_{z}+\Omega_{M}\cos\left(\omega_{M}^{d}t+\phi_{M}\right)\tau_{x},\label{eq:HMeff}
\end{equation}
which has the usual form for a driven two-level system and which we
use to write the effective Hamiltonian $H_{M}^{{\rm eff}}$ for the
mediator dot in Eq. (\ref{eq:Hleff}). 

We also re-express the mediator dot number operators $n_{c1}$ and
$n_{c2}$ in the capacitive interaction $H_{QM}$ {[}Eq. (\ref{eq:HQM})
in the main text{]} using the effective two-level singlet representation.
These number operators are given by 
\begin{align}
n_{c1} & =\ket{S_{12}}\bra{S_{12}}+2\ket{S_{11}}\bra{S_{11}}\nonumber \\
 & =\frac{3}{2}\mathbf{1}-\frac{1}{2}\tau_{z},\nonumber \\
n_{c2} & =\ket{S_{12}}\bra{S_{12}}\nonumber \\
 & =\frac{1}{2}\mathbf{1}+\frac{1}{2}\tau_{z}\label{eq:nc1nc2}
\end{align}
in the singlet basis. Note that $n_{c1}$ and $n_{c2}$ commute with
$U_{{\rm rf}}$ {[}Eq. (\ref{eq:Urf}) in the main text{]} and are
thus invariant under transformation to the rotating frame. We use
the expressions in Eq. (\ref{eq:nc1nc2}) to write the effective interaction
Hamiltonian $H_{QM}^{{\rm eff}}$ in Eq. (\ref{eq:Hleff}).

\section{\label{sec:RXqubitbas} Symmetric resonant exchange qubit}

The RX qubit is an exchange-only qubit \cite{DiVincenzo2000Nature,Laird2010,Gaudreau2011,Medford2013NNano}
defined by the spin states of three electrons in a triple quantum
dot that couples directly to electric fields via the intrinsic spin-charge
mixing present in the logical qubit states \cite{Taylor2013,Medford2013}.
Here, we focus on the case of silicon quantum dots, for which the
three-electron states in the lower-energy $S=1/2$ spin subspace in
the presence of a static magnetic field ${\rm \gtrsim100\ mT}$ \cite{Medford2013}
are those with a spin quantum number for the total $z$ component
$m_{s}=-1/2$ due to the positive g-factor of silicon. An identical
analysis can be carried out for the $\left(S=1/2,m_{s}=1/2\right)$
subspace \cite{Taylor2013}. As in the case of the mediator dot model
discussed in the previous section, we assume a sufficiently large
valley splitting energy $E_{{\rm V}}\gtrsim100\ \mu{\rm eV}$ and
consider only the lowest-energy valley state for each of the three
dots. The $m_{s}=-1/2$ subspace is spanned by the $\left(1,1,1\right)$
states \cite{Taylor2013,Srinivasa2024}

\begin{eqnarray}
\ket{e_{0}} & \equiv & \ket{s}_{13}\ket{\downarrow}_{2}\nonumber \\
 & = & \frac{1}{\sqrt{2}}\left(c_{1\uparrow}^{\dagger}c_{2\downarrow}^{\dagger}c_{3\downarrow}^{\dagger}-c_{1\downarrow}^{\dagger}c_{2\downarrow}^{\dagger}c_{3\uparrow}^{\dagger}\right)\ket{\rm \text{Ø}},\label{eq:state1}\\
\ket{g_{0}} & \equiv & \sqrt{\frac{1}{3}}\ket{t_{0}}_{13}\ket{\downarrow}_{2}-\sqrt{\frac{2}{3}}\ket{t_{-}}_{13}\ket{\uparrow}_{2}\nonumber \\
 & = & \frac{1}{\sqrt{6}}\left(c_{1\uparrow}^{\dagger}c_{2\downarrow}^{\dagger}c_{3\downarrow}^{\dagger}+c_{1\downarrow}^{\dagger}c_{2\downarrow}^{\dagger}c_{3\uparrow}^{\dagger}-2c_{1\downarrow}^{\dagger}c_{2\uparrow}^{\dagger}c_{3\downarrow}^{\dagger}\right)\ket{\rm \text{Ø}},\nonumber \\
\label{eq:state0}
\end{eqnarray}
along with the $\left(2,0,1\right)$ and $\left(1,0,2\right)$ states
\begin{eqnarray}
\ket{s_{1,-1/2}} & \equiv & \ket{s}_{11}\ket{\downarrow}_{3}=c_{1\uparrow}^{\dagger}c_{1\downarrow}^{\dagger}c_{3\downarrow}^{\dagger}\ket{\rm \text{Ø}},\label{eq:stateS11}\\
\ket{s_{3,-1/2}} & \equiv & \ket{\downarrow}_{1}\ket{s}_{33}=c_{1\downarrow}^{\dagger}c_{3\uparrow}^{\dagger}c_{3\downarrow}^{\dagger}\ket{\rm \text{Ø}},\label{eq:stateS33}
\end{eqnarray}
where $\ket{\rm \text{Ø}}$ denotes the vacuum. 

In this work, we consider symmetric operation of each RX qubit $\alpha$
at the fixed operation point $\epsilon_{\alpha}=0$ chosen in the
sideband-based cavity-mediated qubit-qubit entangling approach of
Ref. \cite{Srinivasa2024} to incorporate leading-order protection
from charge noise. As in the case of the sideband-based long-range
entangling gates, the local driven dot-mediated entangling gates we
consider in this work do not require tuning of the qubits away from
the $\epsilon_{\alpha}=0$ operation points, enabling the associated
protection of the qubits from charge noise to be retained during the
two-qubit gate operation. Thus, only single-qubit operations for the
RX qubit, which involve ac driving of the detuning $\epsilon$ \cite{Taylor2013,Medford2013},
require moving to $\epsilon\neq0$ in this approach. For notational
simplicity, we drop the qubit index $\alpha$ elsewhere in this section
unless otherwise specified. 

\begin{figure}
\begin{centering}
\includegraphics[viewport=0bp 0bp 550bp 350bp,width=0.45\textwidth]{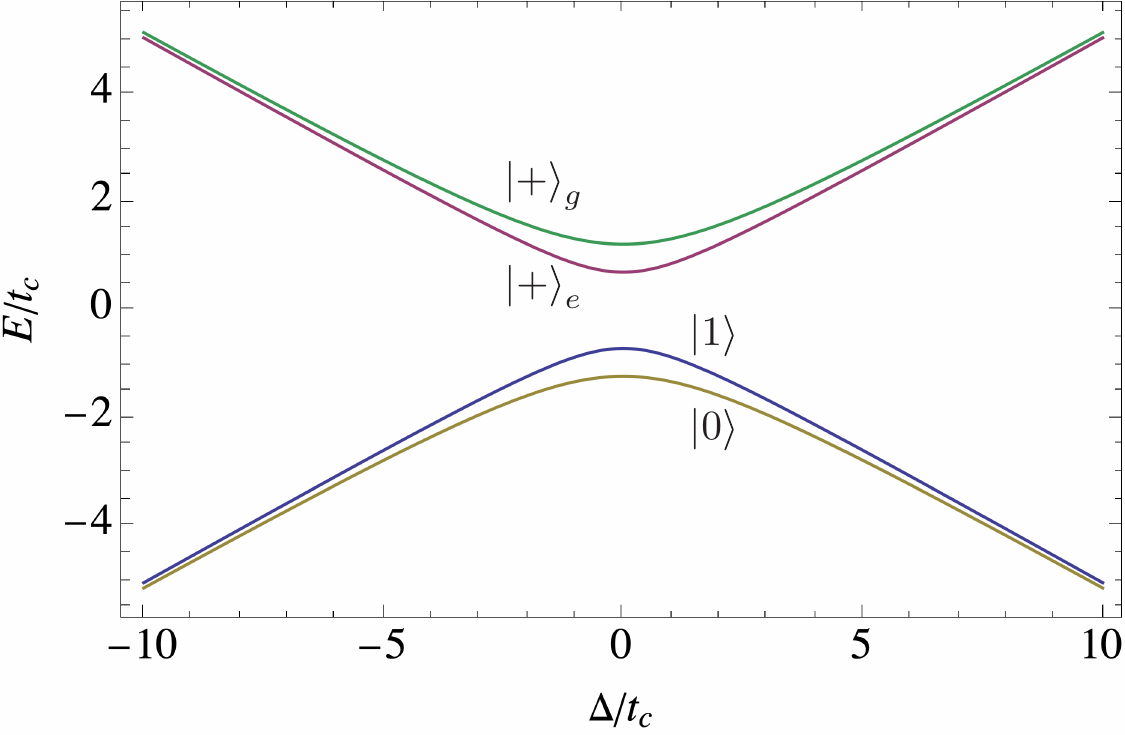}
\par\end{centering}
\caption{\label{fig:spectrum}Spectrum of the symmetric RX qubit Hamiltonian
$H_{{\rm sym}}$ {[}Eq. (\ref{eq:Hsym}){]}, showing the energies
of the four eigenstates of $H_{{\rm sym}}$ given in Eqs. (\ref{eq:pmestates})
and (\ref{eq:pmgstates}) as a function of the half-width $\Delta$
of the $\left(1,1,1\right)$ region of the triple quantum dot, expressed
in units of the tunnel coupling $t_{c}\equiv t_{l}=t_{r}$ for $\epsilon=0$
and with a uniform energy shift $\Delta/2$ dropped. Here, $\ket{0}\equiv\ket{-}_{g}$
and $\ket{1}\equiv\ket{-}_{e}$ represent the logical states of the
undriven RX qubit in the symmetric operation regime. }
\end{figure}

We start from the Hubbard model Hamiltonian for the RX qubit specified
by Eqs. (\ref{eq:HQ})-(\ref{eq:Halphat}) in the basis $\left\{ \ket{e_{0}},\ket{g_{0}},\ket{s_{1,-1/2}},\ket{s_{3,-1/2}}\right\} $
\cite{Taylor2013},
\begin{equation}
H_{{\rm hub}}=\left(\begin{array}{cccc}
0 & 0 & -\frac{t_{l}}{2} & -\frac{t_{r}}{2}\\
0 & 0 & -\frac{\sqrt{3}t_{l}}{2} & \frac{\sqrt{3}t_{r}}{2}\\
-\frac{t_{l}}{2} & -\frac{\sqrt{3}t_{l}}{2} & \Delta+\epsilon & 0\\
-\frac{t_{r}}{2} & \frac{\sqrt{3}t_{r}}{2} & 0 & \Delta-\epsilon
\end{array}\right),\label{eq:Hhub}
\end{equation}
where $\epsilon\equiv-\left(\epsilon_{1}-\epsilon_{3}\right)/2$ represents
the energy detuning between the outer dot orbitals of the triple dot
and $\Delta\equiv U-2V-V_{m}$ sets the width of the $\left(1,1,1\right)$
region in terms of the energy detuning $V_{m}\equiv-\epsilon_{2}+\left(\epsilon_{1}+\epsilon_{3}\right)/2$
between the center dot orbital and the average of the outer dot orbitals
(see Fig. 1 in Ref. \cite{Srinivasa2016}). Setting $\epsilon=0$
and $t_{l}=t_{r}\equiv t_{c},$ we re-express the Hamiltonian in Eq.
(\ref{eq:Hhub}) in the symmetrized basis $\left\{ \ket{e_{0}},\ket{s_{+}},\ket{g_{0}},\ket{s_{-}}\right\} ,$
where $\ket{s_{\pm}}\equiv\left(\ket{s_{1,-1/2}}\pm\ket{s_{3,-1/2}}\right)/\sqrt{2},$
which yields 
\begin{equation}
H_{{\rm sym}}=\left(\begin{array}{cc|cc}
0 & -\frac{t_{c}}{\sqrt{2}}\\
-\frac{t_{c}}{\sqrt{2}} & \Delta\\
\hline  &  & 0 & -\sqrt{\frac{3}{2}}t_{c}\\
 &  & -\sqrt{\frac{3}{2}}t_{c} & \Delta
\end{array}\right).\label{eq:Hsym}
\end{equation}
The block-diagonal structure of $H_{{\rm sym}}$ involves two pairs
of tunnel-coupled states $\left\{ \ket{e_{0}},\ket{s_{+}}\right\} $
and $\left\{ \ket{g_{0}},\ket{s_{-}}\right\} $ and thus enables exact
diagonalization, in contrast to the perturbative derivation of the
effective Hamiltonian for general $\epsilon$ used to describe full
resonant microwave-driven control of the RX qubit \cite{Taylor2013,Medford2013}.
We note that these pairs of coupled states also correspond to quadrupolar
coupling \cite{Friesen2017,Koski2020} for three electrons. Here,
this coupling is responsible for the admixture of the polarized charge
states {[}Eqs. (\ref{eq:stateS11}) and (\ref{eq:stateS33}){]} in
each of the logical RX qubit states. Diagonalizing $H_{{\rm sym}}$
in the subspace $\left\{ \ket{e_{0}},\ket{s_{+}}\right\} $ gives
the eigenstates
\begin{align}
\ket{-}_{e} & =\cos\left(\frac{\theta_{e}}{2}\right)\ket{e_{0}}+\sin\left(\frac{\theta_{e}}{2}\right)\ket{s_{+}},\nonumber \\
\ket{+}_{e} & =-\sin\left(\frac{\theta_{e}}{2}\right)\ket{e_{0}}+\cos\left(\frac{\theta_{e}}{2}\right)\ket{s_{+}},\label{eq:pmestates}
\end{align}
with energies $E_{\pm}^{e}=\Delta/2\pm\Omega_{e}/2,$ where $\Omega_{e}=\sqrt{\Delta^{2}+2t_{c}^{2}}$
and $\tan\theta_{e}=\sqrt{2}t_{c}/\Delta.$ Similarly, diagonalization
in the subspace $\left\{ \ket{g_{0}},\ket{s_{-}}\right\} $ gives
the eigenstates
\begin{align}
\ket{-}_{g} & =\cos\left(\frac{\theta_{g}}{2}\right)\ket{g_{0}}+\sin\left(\frac{\theta_{g}}{2}\right)\ket{s_{-}},\nonumber \\
\ket{+}_{g} & =-\sin\left(\frac{\theta_{g}}{2}\right)\ket{g_{0}}+\cos\left(\frac{\theta_{g}}{2}\right)\ket{s_{-}},\label{eq:pmgstates}
\end{align}
with energies $E_{\pm}^{g}=\Delta/2\pm\Omega_{g}/2,$ where $\Omega_{g}=\sqrt{\Delta^{2}+6t_{c}^{2}}$
and $\tan\theta_{g}=\sqrt{6}t_{c}/\Delta.$ The spectrum is plotted
in Fig. \ref{fig:spectrum} as a function of $\Delta$, where we have
dropped a uniform energy shift $\Delta/2$ in all four energies. 

For the symmetric operation regime considered here, the RX qubit basis
states in the absence of driving are represented by the two lower-energy
eigenstates $\ket{0}\equiv\ket{-}_{g}$ and $\ket{1}\equiv\ket{-}_{e}.$
This approximation is valid provided $\left(\Omega_{g}-\Omega_{e}\right)/2\ll\Omega_{e},$
which is equivalent to the condition $\left[\sqrt{\left(1+6\xi^{2}\right)/\left(1+2\xi^{2}\right)}-1\right]/2\ll1$
with $\xi\equiv t_{c}/\Delta$ denoting the charge admixture parameter
\cite{Taylor2013,Srinivasa2016} and is satisfied for all real $\xi.$
The RX qubit basis states are therefore energetically well separated
from the higher-lying two states, and we can approximate each three-electron
triple dot as an effective two-level system. Accordingly, we write
the RX qubit Hamiltonian {[}Eq. (\ref{eq:HQ}){]} as
\begin{equation}
H_{Q}^{{\rm eff}}\equiv\sum_{\alpha=a,b}\frac{\omega_{\alpha}}{2}\sigma_{\alpha}^{z},\label{eq:HQeff}
\end{equation}
where we have defined $\sigma_{\alpha}^{z}\equiv\ket{1}_{\alpha}\bra{1}-\ket{0}_{\alpha}\bra{0}$
and the RX qubit frequencies $\omega_{\alpha}\equiv\left(\Omega_{\alpha g}-\Omega_{\alpha e}\right)/2.$
The admixture of the polarized $(2,0,1)$ and $\left(1,0,2\right)$
states in the logical qubit basis states $\ket{0}$ and $\ket{1}$
via $\ket{s_{\pm}}$ enables direct interaction of the RX qubit with
electric fields and is essential for the capacitive dot-mediated two-qubit
gate mechanism presented in this work. 

In order to analyze this intramodular entangling approach in terms
consistent with the intermodular coupling approach presented in Ref.
\cite{Srinivasa2024}, which describes cavity photon-mediated entanglement
between distant driven qubits, we transform to the dressed-state basis
for each RX qubit as described in Sec. \ref{subsec:EffHamiltonian}
of the main text. Considering for simplicity the case of resonantly
driven qubits with frequency $\omega$ such that the drive frequency
$\omega^{d}=\omega,$ this transformation is described via the unitary
operator in Eq. (6) of Ref. \cite{Srinivasa2024} with $\theta=\pi/2$
(where we have again dropped the qubit index to simplify the notation).
In the description we use here, the unitary operator for the resonantly
driven case becomes
\begin{equation}
U_{q}=e^{-i\pi\sum_{\alpha}\sigma_{\alpha}^{y}/4},\label{eq:Uq}
\end{equation}
where we now define $\sigma_{\alpha}^{y}\equiv-i\left(\ket{1}_{\alpha}\bra{0}-\ket{0}_{\alpha}\bra{1}\right)$
for $\alpha=a,b.$ Applying the unitary transformation in Eq. (\ref{eq:Uq})
to the Hamiltonian for each RX qubit in the subspace $\left\{ \ket{1}_{\alpha},\ket{0}_{\alpha}\right\} $
{[}Eq. (\ref{eq:HQeff}){]} yields
\begin{align}
\tilde{H}_{Q} & \equiv U_{q}^{\dagger}H_{Q}^{{\rm eff}}U_{q}=-\sum_{\alpha=a,b}\frac{\omega_{\alpha}}{2}\tilde{\sigma}_{\alpha}^{x},
\end{align}
where the Pauli operators in the dressed-state basis are defined according
to $\tilde{\sigma}_{\alpha}^{z}\equiv\ket{e}_{\alpha}\bra{e}-\ket{g}_{\alpha}\bra{g}$
with
\begin{align}
\ket{e} & \equiv\frac{1}{\sqrt{2}}\left(\ket{0}+\ket{1}\right),\nonumber \\
\ket{g} & \equiv\frac{1}{\sqrt{2}}\left(\ket{0}-\ket{1}\right),\label{eq:egdrRXbas}
\end{align}
which specify the dressed states used as the logical RX qubit basis
in Eqs. (\ref{eq:Hq})-(\ref{eq:Uxx}). Note that for the intramodular
entangling approach considered in this work, the RX qubits are not
driven. The transformation to the basis in Eq. (\ref{eq:egdrRXbas})
serves to recast the effective qubit-qubit coupling in a dressed-qubit
representation consistent with that used for the long-distance entangling
approach of Ref. \cite{Srinivasa2024}.

The dependence of the capacitive interaction $H_{QM}$ {[}Eq. (\ref{eq:HQM}){]}
on qubit operators is contained in the occupation number operators
$n_{a3}$ and $n_{b1}$ for the dots in RX qubits $a$ and $b$ that
are adjacent to the mediator dot. In order to express $H_{QM}$ in
the RX qubit basis, we transform $n_{a3}$ and $n_{b1}$ to this basis.
The number operators for the outer dots 1 and 3 of each qubit are
given in the initial basis $\left\{ \ket{e_{0}},\ket{g_{0}},\ket{s_{1,-1/2}},\ket{s_{3,-1/2}}\right\} $
by
\begin{align}
n_{1} & =\left(\begin{array}{cccc}
1 & 0 & 0 & 0\\
0 & 1 & 0 & 0\\
0 & 0 & 2 & 0\\
0 & 0 & 0 & 1
\end{array}\right),\nonumber \\
n_{3} & =\left(\begin{array}{cccc}
1 & 0 & 0 & 0\\
0 & 1 & 0 & 0\\
0 & 0 & 1 & 0\\
0 & 0 & 0 & 2
\end{array}\right).\label{eq:n1n3}
\end{align}
In the symmetrized representation $\left\{ \ket{e_{0}},\ket{s_{+}},\ket{g_{0}},\ket{s_{-}}\right\} $
used for $H_{{\rm sym}}$ in Eq. (\ref{eq:Hsym}), we find 
\begin{equation}
n_{1(3),{\rm sym}}=\left(\begin{array}{cc|cc}
1 & 0 & 0 & 0\\
0 & \frac{3}{2} & 0 & \pm\frac{1}{2}\\
\hline 0 & 0 & 1 & 0\\
0 & \pm\frac{1}{2} & 0 & \frac{3}{2}
\end{array}\right).\label{eq:n1n3sym}
\end{equation}
The triple dot occupation number operators can be used to express
the electric dipole moment of the RX qubit, which can be regarded
as giving rise to the leading-order approximation of the capacitive
interaction \cite{Taylor2013}, as $d_{{\rm RX}}=ew\left(n_{1}-n_{3}\right)/2,$
where $w$ is the triple dot size \cite{Srinivasa2016}. Note that
$d_{{\rm RX}}$ vanishes for the uniform $\left(1,1,1\right)$ triple
dot configuration. In the symmetrized representation of Eq. (\ref{eq:n1n3sym}),
the dipole operator becomes $d_{{\rm RX,sym}}=\frac{ew}{2}\left(\left|s_{+}\right\rangle \left\langle s_{-}\right|+\left|s_{-}\right\rangle \left\langle s_{+}\right|\right).$
Thus, $d_{{\rm RX}}$ is nonzero due to the admixture of the polarized
states $\ket{s_{1,-1/2}}$ and $\ket{s_{3,-1/2}}$ and describes a
transition between the symmetrized states $\ket{s_{\pm}}.$ 

Transforming to the eigenstate basis in Eqs. (\ref{eq:pmestates})
and (\ref{eq:pmgstates}) and applying the effective two-level approximation
used to obtain Eq. (\ref{eq:HQeff}) yields 
\begin{equation}
n_{1\left(3\right),d}^{{\rm eff}}=q_{0}\mathbf{1}-q_{z}\sigma_{z}\pm q_{x}\sigma_{x}\label{eq:n1n3deff}
\end{equation}
in the RX qubit basis, where we have defined 
\begin{align}
q_{0} & \equiv q_{0}\left(\theta_{e},\theta_{g}\right)=\frac{5}{4}-\frac{1}{8}\left(\cos\theta_{e}+\cos\theta_{g}\right),\nonumber \\
q_{z} & \equiv q_{z}\left(\theta_{e},\theta_{g}\right)=\frac{1}{8}\left(\cos\theta_{e}-\cos\theta_{g}\right),\nonumber \\
q_{x} & \equiv q_{x}\left(\theta_{e},\theta_{g}\right)=\frac{1}{2}\sin\left(\frac{\theta_{e}}{2}\right)\sin\left(\frac{\theta_{g}}{2}\right).\label{eq:q0qzqx}
\end{align}
For notational convenience in writing the number operators $n_{b1,d}^{{\rm eff}}$
and $n_{a3,d}^{{\rm eff}}$ appearing in the effective interaction
term $H_{QM}^{{\rm eff}}$ of Eq. (\ref{eq:Hleff}), we use $q_{0}^{\alpha}=q_{0}\left(\theta_{\alpha e},\theta_{\alpha g}\right)$
and $q_{z}^{\alpha}=q_{z}\left(\theta_{\alpha e},\theta_{\alpha g}\right)$
for $\alpha=a,b,$ while we set $q_{x}^{a}=q_{x}\left(\theta_{ae},\theta_{ag}\right)$
and $q_{x}^{b}=-q_{x}\left(\theta_{be},\theta_{bg}\right)$ to take
into account the different signs in front of the $\sigma_{x}$ term
in Eq. (\ref{eq:n1n3deff}). The dimensionless coefficients $q_{0,z,x}^{\alpha}\left(\theta_{\alpha e},\theta_{\alpha g}\right)$
contain the dependence of the dot-mediated capacitive interaction
in Eq. (\ref{eq:Hleff}) on the specific parameters characterizing
the qubits, which are $\left(\Delta_{\alpha},t_{\alpha}\right)$ for
the symmetric RX qubit. As noted in Sec. \ref{subsec:EffHamiltonian},
by replacing these coefficients with those appropriate for other types
of spin qubits that allow for capacitive coupling and incorporating
any required spin-charge mixing mechanisms, the theory presented in
this work can be adapted to a wide range of spin qubit systems. 

\section{\label{sec:RWAvalidity}Validity of rotating wave approximation}

Here, we show that the condition $\Omega_{M}\ll\left|\delta_{\alpha}\right|,\left|\delta_{\alpha}\pm\Omega_{M}\right|$
governing the validity of the rotating wave approximation (RWA) used
to derive the effective Hamiltonian in Sec. \ref{subsec:EffHamiltonian}
is satisfied for typical system parameters. For a resonantly driven
mediator dot such that $\omega_{M}^{d}=\omega_{s}^{\prime}$ and using
Eq. (\ref{eq:omegaprfreqs}), we find
\begin{align}
\left|\delta_{\alpha}\right| & =\left|\omega_{\alpha}^{\prime}-\omega_{s}^{\prime}\right|\nonumber \\
 & =\left|\omega_{\alpha}-\omega_{s}-2q_{z}^{\alpha}K_{0}-\sum_{\alpha}q_{0}^{\alpha}\Delta K\right|.\label{eq:absdeltaalpha}
\end{align}
To gain some intuition for the relative sizes of the frequencies in
$\left|\delta_{\alpha}\right|,$ we first note that for the mediator
dot, $\omega_{s}=\Delta_{T}+2J_{c}$ {[}see Fig. \ref{fig:mediatordotspectrum}
and Appendix \ref{sec:MEQDdipole}{]}. On the other hand, for symmetric
resonant exchange qubits ($\epsilon_{\alpha}=0$ and $t_{\alpha l}=t_{\alpha r}\equiv t_{\alpha}$),
the qubit frequencies derived from the perturbative model are $\omega_{\alpha}=J_{\alpha l}=J_{\alpha r}=J_{\alpha}=t_{\alpha}^{2}/\Delta_{\alpha}$
for $\alpha=a,b,$ where $J_{\alpha}$ is the interdot exchange interaction
strength \cite{Taylor2013,Srinivasa2016}. In quantum dot systems,
the on-site exchange interaction $2J_{c}$ is typically much larger
than the interdot exchange interaction \cite{Malinowski2019}, so
that $\omega_{s}>2J_{c}\gg J_{\alpha}=\omega_{\alpha}$ and $\left|\delta_{\alpha}\right|=\omega_{s}+2q_{z}^{\alpha}K_{0}+\sum_{\alpha}q_{0}^{\alpha}\Delta K-\omega_{\alpha}$
is a large positive quantity. 

Using the exact expression for the symmetric RX qubit frequency $\omega_{\alpha}\equiv\left(\Omega_{\alpha g}-\Omega_{\alpha e}\right)/2=\left(\sqrt{\Delta_{\alpha}^{2}+6t_{\alpha}^{2}}-\sqrt{\Delta_{\alpha}^{2}+2t_{\alpha}^{2}}\right)/2$
found in Appendix \ref{sec:RXqubitbas} along with Eq. (\ref{eq:q0qzqx})
and the parameters chosen in the present work (see Sec. \ref{subsec:parameters_Kab}),
we find $\omega_{a}=\omega_{b}\equiv\omega\approx2\pi\times4.9\ {\rm GHz},$
$2q_{z}^{a}K_{0}=2q_{z}^{b}K_{0}\approx2\pi\times5.8\ {\rm GHz},$
and $\left(q_{0}^{a}+q_{0}^{b}\right)\Delta K\approx2\pi\times10\ {\rm GHz}.$
These values, together with the constraint $\delta_{a}=\delta_{b}=8r\mathcal{K}_{ab}$
for the two-qubit gate $U_{xx}$ in Eq. (\ref{eq:Uxx}) and the choice
$r=-100,$ give $\omega_{s}\approx2\pi\times16\ {\rm GHz}$ and $\left|\delta_{a}\right|=\left|\delta_{b}\right|\approx2\pi\times27\ {\rm GHz.}$
As this frequency far exceeds the mediator dot drive Rabi frequency
$\Omega_{M}\approx2\pi\times97\ {\rm MHz},$ the condition $\Omega_{M}\ll\left|\delta_{\alpha}\right|,\left|\delta_{\alpha}\pm\Omega_{M}\right|$
for the RWA is satisfied. This parameter regime, which is enabled
by the typical case $2J_{c}\gg J_{\alpha}$ and the positive energy
shifts from the capacitive coupling, leads to a separation of energy
scales that simplifies the form of the effective interaction mediated
by the driven dot {[}Eq. (\ref{eq:HleffRWA}){]} as shown in Sec.
\ref{subsec:EffHamiltonian}.

\bibliography{dotmedRXbib}

\end{document}